\documentclass[journal]{IEEEtran}
\pdfoutput=1
\usepackage{color} 
\usepackage{xcolor,colortbl}
\definecolor{DarkRed}{rgb}{0.8,0,0}
\definecolor{Blue}{rgb}{0,0,0.8}
\definecolor{Gray}{rgb}{0.9,0.9,0.9}

\usepackage{amsmath,amsfonts,amssymb}
\usepackage{amsthm}
\allowdisplaybreaks
\usepackage{mathrsfs} 
\usepackage{units}
\usepackage{multirow}
\usepackage{tabularx}

\usepackage{lscape}

\usepackage{graphicx} 
\usepackage[tight,footnotesize]{subfigure} 
\usepackage{array} 
\usepackage[footnotesize]{caption} 
\usepackage{tabularx}
\usepackage{ctable}
\usepackage{booktabs}

\usepackage{url}

\usepackage{breakurl}

\DeclareMathOperator{\oD}{D}                                              
\newcommand{\oT}{\textnormal{T}}                                   
\newcommand{\oone}{\mathbf{1}}                                            

\newtheorem{proposition}{Proposition}
\newtheorem{definition}[proposition]{Definition}

\newtheorem{example}[proposition]{Example}

\begin{document}

\title{\emph{MyAdChoices}: Bringing Transparency and Control to Online Advertising}

\author{Javier Parra-Arnau and Jagdish Prasad Achara and Claude Castelluccia 
\IEEEcompsocitemizethanks{\IEEEcompsocthanksitem The authors are with
the Privatics research team, INRIA Grenoble - Rh\^one-Alpes (France),
\protect\\
E-mail: xparnau@gmail.com, \{jagdish.achara, claude.castelluccia\}@inria.fr}
\thanks{Manuscript prepared February, 2016.}}

\IEEEcompsoctitleabstractindextext{%
\begin{abstract}
The intrusiveness and the increasing invasiveness of online advertising have, in the last few years, raised serious concerns regarding user privacy and Web usability. As a reaction to these concerns, we have witnessed the emergence of a myriad of ad-blocking and anti-tracking tools, whose aim is to return control to users over advertising. The problem with these technologies, however, is that they are extremely limited and radical in their approach: users can only choose either to block or allow all ads. With around 200 million people regularly using these tools, the economic model of the Web ---in which users get content free in return for allowing advertisers to show them ads--- is at serious peril.
In this paper, we propose a smart Web technology that aims at bringing transparency to online advertising, so that users can make an informed and equitable decision regarding ad blocking. The proposed technology is implemented as a Web-browser extension and enables users to exert fine-grained control over advertising,
thus providing them with certain guarantees in terms of privacy and browsing experience, while preserving the Internet economic model.
Experimental results in a real environment demonstrate the suitability and feasibility of our approach,
and provide preliminary findings on behavioral targeting from real user browsing profiles.
\end{abstract}

}

\maketitle

\IEEEdisplaynotcompsoctitleabstractindextext

%
\IEEEpeerreviewmaketitle

\setcounter{footnote}{0} 

\section{Introduction}\label{sec:intro}
\noindent
During the last two decades, the Internet and the World Wide Web have been gradually integrating into people's daily lives, enabling new forms of communication such as e-mail and instant messaging.
The so-called network of networks has become an essential communication channel not only among people, but also among businesses and their customers.

Breathing new life into traditional business activities is precisely one of the Internet's most relevant influences.
The Web has led to key business changes embracing the whole value chain in almost all sectors and companies.
These changes have had an impact on how products are sold and also, and more importantly, on how companies approach customers in a personalized manner, taking into account their unique preferences.

The industry of advertising,
lavishly illustrated by Yahoo! Advertising, Google DoubleClick and real-time bidding (RTB),
is a clear example of the transformation driven by the ever-growing sophistication of Web technologies.
In the past, ads were served directly by
the Web site's owner
following a one-size-fits-all approach.
But due to the gradual introduction
of intermediary companies with extensive capabilities to track users,
Internet advertising has become increasingly personalized and pervasive.

The ability of the online marketing industry to track and profile users' Web-browsing activity
is therefore what enables more effective, tailored-made advertising services.
The intrusiveness of these practices and the increasing invasiveness of digital advertising, however,
have raised serious concerns regarding user privacy and Web usability.
According to recent surveys,
two out of three Internet users are worried about the fact that their online behavior be scrutinized without their knowledge and consent~\cite{Purcell12PIALP}.
Numerous studies in this same line reflect the growing level of ubiquity and abuse of advertising,
which is perceived by users as a significant degradation of their browsing experience~\cite{Adobe12TR,Marvin13TR,PageFair15AdBlock}.

In response to these concerns, recent years have witnessed the rise of a myriad of ad-blocking
tools whose primary aim is to return control to users over advertising.
In essence, ad blockers monitor all network connections that may be initiated when the browser loads a page,
and prevent those which are made with third parties\footnote{These connections are often referred to as \emph{third-party network requests}, while those established with the page's owner are called \emph{first-party network requests}.} and may correspond to ads.
To this end, ad blockers rely on blacklists manually maintained by their developing companies and, in some cases, by user communities.

Apart from the controversy stirred by the use of such lists ---especially after the revelation that Adblock Plus~\cite{AdblockPlus15Web}, the most popular of these technologies, was getting money from ad companies to whitelist them~\cite{Cookson15FT}---,
the main problem with these tools is that they were conceived without
considering two key points:
first, the crucial role of online advertising as the major sustainer of the Internet ``free'' services;
and secondly, the social and economic benefit of non-intrusive and rational advertising.
While ad-blockers might constitute a first attempt in this bid to regain control over advertising,
they are extremely limited and radical in their approach: users can only choose either to block or allow all the ads blacklisted by the ad-blocking companies.

In a half-hearted attempt to address the aforementioned privacy and usability concerns, the Internet advertising industry and the World Wide Web Consortium have participated in two self-regulatory initiatives, Your Online Choices~\cite{YourOnlineChoicesWeb} and Do Not Track (DNT)~\cite{DNT15W3C}.
Although these two initiatives make opt-out easier for users  ---the former to stop receiving ads tailored to their Web-browsing interests,
and the latter to stop being tracked through third-party cookies---,
the fact is that users have no control over whether or not their advertising and tracking preferences are honored.

With around 200 million people worldwide regularly using ad blockers\footnote{Adblock Plus is Google Chrome's most popular plug-in the world with more than 50 million monthly active users, and an increase of 41 percent in the last year.}, as well as with Apple's recent support for the development of such tools in
its new iOS release~\cite{Naughton15GUARDIAN},
the economic model underlying the Web is at serious risk~\cite{PageFair15AdBlock}.
This has spurred a heated debate about the ethics of these technologies and the need for a solution that strikes a better balance among the Internet's dominant business model, user privacy and Web usability~\cite{Arment15Blog,David15MediaPost,Thielman15GUARDIAN}.

We believe that the solution necessarily implies giving users \emph{real} control over advertising,
and that this can only be achieved through technologies that enforce their actual preferences, and not the radical, binary choices provided by the current
ad blockers.
As a matter of fact, according to a recent survey,
two out of three ad-blocker users are not against ads
and would accept the trade-off that comes with the ``free'' content~\cite{AdblockPlus15Survey};
this is provided that advertising is a transparent process and they have control over the personal information that is collected~\cite{Rogers15Forbes}.
Trust, through transparency, seems to be key in this regard~\cite{Morey15HAR}.
However, because different users may have different motivations, we require tools that allow for such different choices regarding ad blocking.

\subsection{Contribution and Plan of this Paper}
\label{sec:intro:contribution}
\noindent
In this work, we investigate a smart Web technology that can bring transparency to online advertising
and help users enforce their own choices over ads.
The technology proposed in this paper has been contrived within the project \emph{MyRealOnlineChoices}, and aims at providing
ad transparency on the one hand, and ad-blocking functionalities on the other.

The main goal of this tool is, first, to let users know what is happening behind the scenes with their Web-browsing data;
and secondly, to enable them to react accordingly, in a flexible and non-radical way, by giving them fine-grained control over advertising.
Its ultimate aim is to provide users with certain guarantees in terms of privacy and browsing experience,
while preserving the online publishing's dominant business model.

Next, we summarize the major contributions of this work:
\begin{itemize}

\item We propose a theoretical model for the investigation of \emph{behavioral targeting},
 a widespread form of advertising that uses information gathered from users' Web-browsing behavior to serve them ads.
The proposed model aims at providing transparency to this ad-serving process. First, by detecting such form of ad-targeting and thus quantifying the extent to which user-browsing interests are exploited. And secondly, by examining the uniqueness of the browsing profiles compiled by the entities that participate in said process.

The strength of the proposed model lies in its more general and mathematically grounded approach to the problem of detecting such form of advertising.
This is unlike previous work which relies on basic heuristics and extremely limiting assumptions, or which oversimplifies the ad-delivery process.
    The detection of behavioral advertising is, in this work, formulated as an optimization problem that reflects the uncertainty in determining the information available at ad platforms and trackers.
    The proposed model capitalizes on fundamental results from the fields of statistical estimation and robust optimization,
    the latter being a relatively new approach to optimization problems affected by uncertainty, but which has already proved useful in applications like signal processing, communication networks and portfolio optimization.

 \item In this same line of transparency and taking this model a step further, we propose a second detection system that sheds light on the uniqueness of the browsing profiles compiled by the entities that participate in the ad-delivery process. To this end, we adopt a quantifiable measure of user-profile uniqueness---the Kullback-Leibler (KL) divergence or \emph{relative entropy} between the probability distribution of the user's Web-browsing interests and the population's distribution, a quantity that we justified and interpreted in~\cite{Parra13FGCS, Rebollo11SECTECH} by leveraging on the rationale behind entropy-maximization methods.

  \item We design a system architecture that
  implements the two aforementioned detection systems as main transparency factors,
  and enables smart ad blocking through the specification of user-configurable control policies.
      The system is designed to provide ad transparency and blocking services all in real-time,
      without the need of any external entity, and by relying on local Web-content categorization and open-source optimization libraries.
      The only exception is the computation of the profile uniqueness, which requires the involvement of an external server.
      A relevant aspect of our system is that it has been conceived to work under two distinct scenarios in terms of tracking,
      which allows users to configure the ad-transparency functionality according to their own perceptions in this respect.
The proposed system architecture is developed in the form of a Web-browser extension for Google Chrome,
and its beta version is available under request.

\item We conduct an experimental analysis from the user data collected by this extension.
Such analysis allows us, first, to evaluate the proposed system in a real environment; and secondly, to investigate several aspects related to behavioral advertising.
The conducted experiments constitute the first attempt to study behavioral targeting from real user browsing profiles.
\end{itemize}

The remainder of this work is organized as follows.
Sec.~\ref{sec:Background}
provides the necessary background in online advertising.
Then, Sec.~\ref{sec:System} presents the theoretical model for the detection of interest-based ads and profile uniqueness.
Sec.~\ref{sec:Implementation} describes the main components of a system architecture that aims at providing ad transparency and advanced ad-blocking functionalities.
Sec.~\ref{sec:Evaluation} analyzes the data collected by the proposed tool in an experiment with 40 participants.
Sec.~\ref{sec:RelatedWork} reviews the state of the art relevant to this work.
Conclusions are drawn in Sec.~\ref{sec:Conclusion}.
Finally, Appendices~\ref{Appendix:LP} and~\ref{Appendix:Feasibility} show, respectively, the linear-program formulation of the interest-based ad detector, and the feasibility of this optimization problem.

\section{Background in Online Advertising}
\label{sec:Background}
\noindent
This section examines the online advertising ecosystem, providing the reader with the necessary depth to understand the technical contributions of this work.
First, Sec.~\ref{sec:Background:Actors} gives an overview of the main actors of this ecosystem. Afterwards, Sec.~\ref{sec:Background:AdServing} describes how ads are served on the Web, and then, Sec.~\ref{sec:Background:AdClasses} provides a standard classification of the targeting objectives commonly available to advertisers. Finally, Sec.~\ref{sec:Background:RTB} presents one of the technologies enabling this ad-serving process. For a detailed, complete explanation on the subject, the reader is referred to~\cite{Smith14B}.

\subsection{Key Actors}
\label{sec:Background:Actors}
\noindent
The online advertising industry is composed by a considerable number of entities with very specific and complementary roles, whose ultimate aim is to display ads on Web sites. Publishers, advertisers, ad platforms, ad agencies, aggregators and optimizers are some of the parties involved in the ad-delivery process~\cite{Yuan12arXiv}.
Despite the enormous complexity\footnote{ The intricacy of the advertising ecosystem is often illustrated in conferences and related venues with the diagram available at~\cite{KawajaLUMAscape}.} and constant evolution of the advertising ecosystem, the process whereby ads are presented on Web sites
is usually characterized or modeled in terms of publishers, advertisers and ad platforms~\cite{Toubiana07S,LiuHotNet13,YanWWW09,AlyWWW12,Tsang04EC}.
Next, we provide a description of these three key actors:

\begin{itemize}
  \item A \emph{publisher} is an entity that owns a Web page (or a Web site) and that, in exchange of some economic compensation, is willing to place ads of other parties in some spaces of its page (or site). An example of publisher is The New York Times' Web site.

  \item An \emph{advertiser} is an entity that wants to display ads on one of the spaces offered by a publisher, and is disposed to pay for it.
      Advertisers typically engage the services of one or several \emph{ad platforms} (described below), which are the ones responsible for displaying their ads on the publishers' sites.
       As we shall explain later in Sec.\ref{sec:Background:AdServing}, there exist two ad-platform models, allowing users to have two different roles.
      In the traditional albeit prevailing approach, advertisers
      indicate the targeting objective/s most suitable for their campaigns, that is, to which users they want their ads to be shown. For example, an advertiser may want the ad platform to serve its ads to an audience interested in politics or to people living in France.
      Advertisers must also specify the amount of money they are willing to pay each time their ads are displayed, and each time users click on them\footnote{In the terminology of online advertising, these quantities are referred to as the cost-per-impression (CPI) and the cost-per-click (CPC), respectively.}.
      On the contrary, in the recently established model of \emph{real-time bidding} (RTB), ad platforms allow advertisers to bid for each ad-impression at the time the user's browser loads a page.
      This model enables advertisers to make their own decisions rather than relying on an intermediary to make decisions for them~\cite{Smith14B}.

  \item An \emph{advertising platform} or \emph{ad platform} is a group of entities that connects advertisers to publishers, i.e., it receives ads from advertisers and places them on the spaces available at publishers. To this end,
      ad platforms track and profile users with the aim of targeting ads to their interests, location and other personal data.
      As we shall describe in greater detail in the next subsection, traditional ad platforms carry out this targeting on their own, in accordance with the campaign requirements and objectives specified by advertisers.
      RTB-based ad platforms, on the other hand,
      share certain user-tracking data with advertisers, which then take charge of selecting who suits them by deciding which user to bid for.
      Some examples of ad platforms include DoubleClick, Gemini and Bing Ads, owned respectively by Google, Yahoo! and Microsoft.
\end{itemize}

\subsection{Ad-Serving Process}
\label{sec:Background:AdServing}
\noindent
Without loss of rigor, throughout this work we shall assume an online advertising model composed mainly of the three entities set forth in the previous subsection.
In this simplified albeit comprehensive terms,
the ad-delivery process begins with publishers embedding in their sites a link to the ad platform/s they want to work with.
The upshot is as follows: when a user retrieves one of those Web sites and loads it, their browser is immediately directed to all the embedded links.
Then, through the use of third-party cookies, Web fingerprinting or other tracking technologies, the ad platform is able to track the user's visit to this and any other site partnering with it.

As one might guess, the ability of tracking users across the Web is of paramount
importance for ad platforms: it enables them to learn the Web page being visited and hence its content; the user's location through their IP address; and, more importantly, their Web-browsing interests.
Afterwards, all these invaluable data about the user is what allow ad platforms to serve \emph{targeted} ads.

To carry out this task,
the vast majority of ad platforms rely on proprietary \emph{targeting algorithms}~\cite{Smith14B}.
The aforementioned user data and the objectives and budgets of all advertisers for displaying their ads are the inputs of these algorithms, which are responsible for selecting which ad will be shown in a particular ad space.
Evidently, their primary aim is to maximize ad-platforms' revenues whilst satisfying advertisers' demand.

As anticipated in Sec.~\ref{sec:Background:Actors}, a new class of ad platforms has recently emerged that
delegates
this targeting process to external third parties,
which then compete in real-time auctions for the impression of their ads.
Ad platforms relying on this scheme usually share information about the user with these parties
so that they can decide whether to bid or not for an ad-impression.
Typically, the entities participating in these auctions are big advertising
agencies representing small and medium advertisers\footnote{A special class of these agencies are the \emph{demand-side platforms} (DSPs), which are systems that automate the purchasing of online advertising on behalf of advertisers.},
and traditional ad platforms wishing to sell the remnant inventory.
This ad-serving scheme is called RTB
and its major advantage, compared to the traditional ad platforms, is to enable advertisers (or others acting on their behalf) to buy individual impressions without having to rely on the ad platform's targeting decision.
In other words, advertisers can decide
whether a particular user is the right person to whom to present their ads.

Finally, regardless of the type of ad platform involved (i.e., RTB-based or not), the ad-serving process ends up by displaying the selected ad in the user's Web browser, a last step that may entail a content-delivery network.

Last but not least,
we would like to stress that the advertising model described here ---and considered in this work--- corresponds to \emph{indirect-sale advertising}, also called network-based or third-party advertising. This is in contrast to the \emph{direct-sale advertisement} model, where publishers and advertisers negotiate directly, without the mediation of ad platforms. In this latter case, we mostly find popular Web sites selling ad space directly to large advertisers. The ads served this way are essentially untargeted, and are often displayed in Web sites where the products and services advertised are related to their contents.
This is mainly because the capability of a publisher to track and profile users is limited just to its own Web site and maybe a few partners. For example, the New York Times' Web site may track users also across the International Herald Tribune, owned by its media group. Such a tracking capability, however, is ridiculous when we compare it with the 2 million sites reachable by Google's ad platforms~\cite{AudienceGuide11}.

\subsection{User-Targeting Objectives}
\label{sec:Background:AdClasses}
\noindent
The ads delivered through indirect-sale advertising allow advertisers to target different aspects of a Web user.
The most popular \emph{targeting objectives} include serving ads tailored to the Web page they are currently visiting, their geographic location, and their Web-browsing interests. Depending on the objective chosen by an advertiser,
ads are classified accordingly as \emph{contextual}, \emph{location-based}, \emph{interest-based} and \emph{untargeted} ads.
Occasionally, we shall refer to these four type of ads as \emph{ad classes}.
Next, we briefly elaborate on each them.

\begin{itemize}
  \item \emph{Contextual ads}. Advertisers can reach their audience through contextual and semantic advertising, by directing ads related to the content of the Web site where they are to be displayed.
      An example of such targeting strategy would be a health-insurance company wishing to show their ads in Web sites whose content is classified as ``health \& fitness''.

  \item \emph{Location-based ads}. They are generated based on the user's location, for example,
  given by the GPS of their smartphone or tablet,
  and also according to the Wi-Fi access points and IP address of the user's machine or device.
  Geographically-targeted ads enable advertisers to launch campaigns targeting users within a certain geographical location.
      An example would be the advertisement of a local music event to users reporting nearby locations.

  \item \emph{Interest-based} or \emph{profile-based ads}. Advertisers can also target users based on their Web-browsing interests. Usually, such interests are inferred from the pages tracked by ad platforms and other tracking companies that may share this information with the former.
    The sequence of Web sites browsed by a user and effectively tracked by an ad-platform or tracker is referred to as the user's \emph{clickstream}.
  In current practice, this is the information leveraged by the online advertising industry to construct a user's interest profile~\cite{Clickstream,Toubiana10SNDSS,CiscoGuide,LiuHotNet13,YanWWW09,AlyWWW12,PandeyCIKM11, Smith14B}.

  \item \emph{Generic ads}. Advertisers can also specify ad placements or sections of publisher's Web sites (among those partnering with the ad platform) where their ads will be displayed.
      Ads served through placement targeting are not necessarily in line with the Web site's content, but may simply respond to some match between the interests of the visiting users and the products advertised.
      Because these ads do not rely on any user data, we shall also refer to them as \emph{generic ads}.
\end{itemize}

An important aspect of the ad-classes described above is that the former three are not mutually exclusive. In other words, except for placement ads ---which are considered to be untargeted---, ads can be simultaneously directed based on content, location and interests. Accordingly, when we refer to interest-based ads, we shall mean that they are targeted \emph{at least} to browsing-interests data. We shall refer to content- and location-based ads in an analogous manner.

In the terminology of online advertising, directing interest-based ads is often called \emph{behavioral targeting}. Another quite popular ad-targeting strategy is \emph{retargeting}, which helps advertisers reach users who previously visited their Web sites. For example, after having browsed Apple's Web site, a user could be shown ads about a new iPhone release when visiting other sites, in an attempt to bring them back.

\begin{figure}
\centering
\includegraphics[width=\columnwidth]{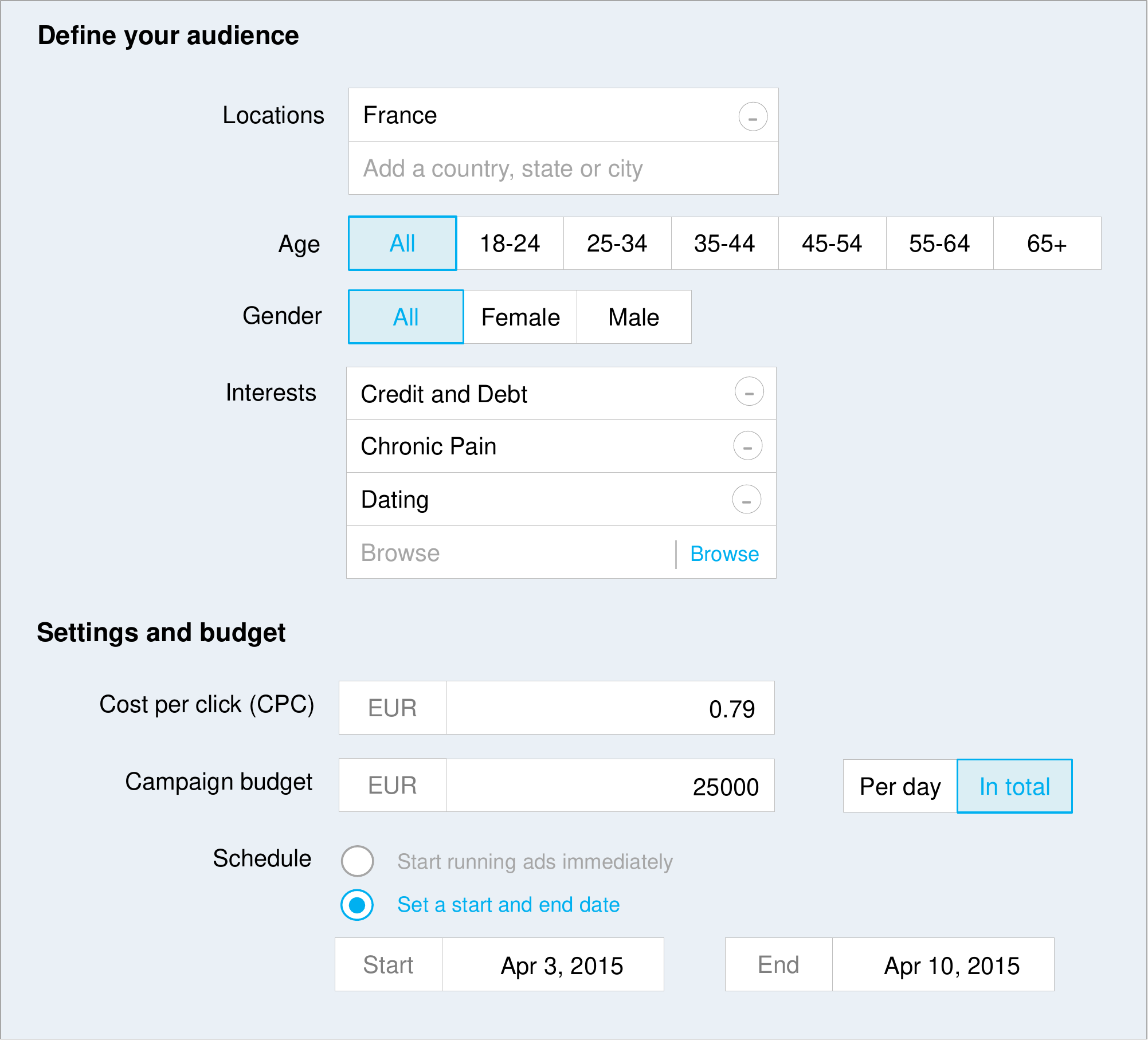}
\caption{Gemini, Yahoo!'s ad platform, offers advertisers the possibility to target ads based on a number of parameters, including the user's browsing interests, which are chosen from a predefined set of 281 bottom-level categories. The categories selected in this example merely show the sensitive, personal information involved in these transactions, and thus do not reflect a real marketing campaign.}
\label{fig:ExampleTargeting}
\end{figure}

We conclude this subsection by giving a real-world example of how advertisers can target their ads.
Fig.~\ref{fig:ExampleTargeting} shows the configuration panel available at Yahoo!'s ad platform, whereby advertisers can define their target audiences based on location, age, gender, interests\footnote{Others platforms like Google's allow advertisers to specify further constraints such as the time of the day ads will be shown, their frequency of appearance to a same user and specific ad-placements.} and context (not shown in this figure). For each campaign, the advertiser must configure all these variables appropriately, evidently with a constraint on the advertising budget.

\subsection{Cookie Matching and Real-Time Bidding}
\label{sec:Background:RTB}
\noindent
This last subsection explains in greater detail some key operational aspects of RTB, an ad-serving scheme that accounts for 20\% of digital ad sales~\cite{Smith14B} but that is expected to be the dominant advertising paradigm in the next years~\cite{eMarketer14a}.

In Sec.~\ref{sec:Background:AdServing}, we mentioned that RTB-based ad platforms share user information
with certain entities,
which then may bid for the impression of their ads.
The auction participants typically include
agencies representing advertisers, DSPs and traditional ad platforms.
To facilitate the sharing of information with these bidders, RTB relies on a \emph{cookie-matching} protocol.

Generally speaking, cookie matching is a process by which two different domains link the user IDs that they have assigned to a same user and that they store in their respective cookies.
Typically, the process is conducted as follows.
When a user visits the former domain,
this domain redirects their browser to the latter domain, including its user ID as a parameter in the URL.
Then, upon receiving the request, the latter domain links this ID with its own ID for this user~\cite{Englehardt14FreeTinker}.

Cookie matching finds its most common application in RTB, where
it allows the ad platform and the bidder to match their cookies for a particular user~\cite{CookieMatchingDC}.
Usually, the protocol is executed only if the bidder wins an auction and delivers its ad to this particular user.
The matching permits the bidder to look up the user (if present) in its own database.
Also,
if subsequent ad-auctions are hold for this user, the bidder will learn that the user information provided in those auctions refer to this same matched user.
We must emphasize that this is under the assumption that this bidder is among the recipients of the bid requests sent by the ad platform.

Having described the technology underlying RTB, next we briefly examine the overall functioning of Google's scheme, probably the most representative.
The following, however, is also valid for other RTB-based ad platforms, although with slight variations irrelevant to this work.

When a user visits a Web site with an ad space served through RTB, an HTTP request is submitted to the ad platform,
which subsequently sends \emph{bid requests} to potential participants.
We note that the number and type of participants involved may vary on a per-auction basis, at the ad platform's discretion.
Within the bid request, the ad platform
generally includes the following data: the URL of the page being visited by the user; the topic category of the page; the user's IP address or parts of it; and other information related to their Web browser~\cite{CookieMatchingRequestDC}.
Accompanying this information, Google's ad platform incorporates a bidder-specific user ID,
which implies that different bidders are given different IDs for a same user.
Other RTB-based ad platforms, alternatively, include their own user's cookies.

Upon receiving the bid request, the bidder may identify the user within its own database through the cookie or identifier. This is provided that the cookie-matching protocol has been executed previously for this user.
Thanks to such cookie or identifier, the bidder can track them across those Web pages in which it is invited to bid.
From those tracked pages, the bidder can therefore build a profile\footnote{DoubleClick's guideline specifies that, unless a bidder wins a given impression, it must not use the data for that impression to profile users~\cite{DoubleClickGuidelines}. Nevertheless, because no active mechanism is enabled to enforce this, nothing prevents a bidder from misusing such user data.}, maybe complementing tracking and other personal data it may have about the user.

The bid price is then set on the basis of the bidder's targeting objectives, that is, whether it aims to target users visiting certain site categories, browsing from a given location, and/or having some specific profile.
To evaluate if the ad-impression meets such objectives, the bidder relies on the aforementioned profile and the information included in the bid request.
If interested, the bidder submits a price to the ad platform, which finally, in a last step, allows the winning bidder to deliver the ad to the user.
It is worth stressing that all this process of gathering user data, ad bidding and delivering is conducted in just tens of milliseconds.

\section{Detection of Profile-based Ad-Serving and Profile Uniqueness}
\label{sec:System}
\noindent
As described in the background section, ad platforms, tracking companies and also advertisers gather information about users (e.g., the visited pages and their location) while they browse the Web. Later, these and other data are leveraged to present ads targeted to the content of the pages browsed, their current geographic location and/or their interests. We also mentioned that ad platforms may as well deliver placement ads, which are considered generic or untargeted ads.

This section investigates a mathematical model that aims at quantifying to what extent the information gathered about a user's \emph{browsing interests} is exploited afterwards by the online advertising industry to serve them ads.
The proposed model
focuses on the detection of interest-based ads since they are the result, and probably the cause, of tracking and profiling users' browsing habits throughout the Internet, often without their knowledge~\cite{Olejnik15PhD} and consent\footnote{Consistently with the recommendations of the US Federal Trade Commission, the advertising industry has started to offer an opt-out scheme for behavioral advertising~\cite{NAI15Opt-out}.}.
It is important to remark that the conducted analysis is restricted to network-based advertisement, as the capability of publishers to track and profile users is, in general, limited to their sites.

In addition to determining if the displayed ads may have been targeted to a browsing profile,
this section addresses another inescapable question related to profile targeting: \emph{how unique} are we seen through the eyes of the companies displaying ads to~us?
As we shall elaborate on in Sec.~\ref{sec:System:DetectionUnique}, the risk of profiling as well as the uniqueness of the profiles built by these companies is closely linked to the risk of reidentification.

In the coming sections, we shall provide the conceptual basis and fundamental operational structure of two detectors that aim at (1) identifying profile-based ads from their interest categories; and (2) shedding light on the uniqueness of the profiles compiled by the entities that participate in the ad-delivery process.
In doing so, we make a preliminary step towards studying the commercial relevance of our browsing history
and quantifying its actual impact on user privacy.
Later in Sec.~\ref{sec:Implementation} we shall present \emph{MyAdChoices}, a Web-browser extension that capitalizes on these two detectors to bring transparency into said process and to enable selective and smart ad-blocking.

\subsection{Statistical and Information-Theoretic Preliminaries}
\label{sec:System:Preliminaries}
\noindent
This section establishes notational aspects and recalls a key information-theoretic concept assumed to be known in the remainder of this paper.

The measurable space in which a \emph{random variable} (r.v.)~takes on values will be called an \emph{alphabet}.
Without loss of generality, we shall always assume that the alphabet is discrete.
We shall follow the convention of using uppercase letters for r.v.'s, and lowercase letters for particular values they take on.
The probability mass function (PMF)~$p$
of an r.v.~$X$ is a function that maps the values taken by $X$ to their probabilities.
Conceptually, a PMF is a \emph{relative histogram} across the possible values determined by its alphabet.

Throughout this work, PMFs will be subindexed by their corresponding r.v.'s in case of ambiguity risk.
Accordingly, both $p(x)$ and $p_X(x)$ denote the value of the function $p_X$ at~$x$.
Occasionally, we shall refer to the function $p$ by its value $p(x)$.
We use the notations $p_{X|Y}$ and $p(x|y)$ equivalently.

We adopt the same notation for information-theoretic quantities used in~\cite{Cover06B}.
Concordantly, the symbol~$\oD$ will denote relative entropy or  KL divergence.
We briefly recall this concept for the reader not intimately familiar with information theory.
All logarithms are taken to base~2.

Given two probability distributions $p(x)$ and $q(x)$ over the same alphabet, the \emph{KL divergence}
$\oD(p\,\|\,q)$ is defined as
$$\oD(p\,\|\,q) = \sum_x p(x) \log \frac{p(x)}{q(x)}.$$
The KL divergence is often referred to as \emph{relative entropy}, as it may be regarded as a
generalization of the Shannon's entropy of a distribution, relative to another.

Although the KL  divergence is not a distance in the mathematical sense of the term,
because it is neither symmetric nor satisfies the triangle inequality,
it does provide a measure of discrepancy between distributions,
in the sense that $\oD(p\,\|\,q)\geqslant0$, with equality if, and only if, $p=q$.

\subsection{Detection of Profile-based Ads}
\label{sec:System:DetectionIB}
\noindent
One of the key functionalities of our system is the detection of profile-based ads, that is, ads that are tailored to a user's browsing interests and, in addition but not necessarily, to their location and the Web-page currently visited.
This section proposes a mathematical model for the identification of these ads, which leverages fundamental results from statistical estimation and robust optimization.

\subsubsection{Ad-Serving Interest-Category Model}
\label{sec:System:DetectionIB:Model}
\noindent
We model the ads delivered by an ad platform (RTB-based or not) to a particular user as independent r.v.'s taking on values on a common finite alphabet of categories or topics, namely the set $\mathscr{X}=\{1,\ldots,n\}$ for some integer $n > 1$.
We hasten to stress that our model encompasses the four classes of ads, or objectives, described in Sec.~\ref{sec:Background:AdClasses}.
The fact that each ad is associated with an interest category does not mean we are considering just interest-based ads. For example, a content-based ad displayed on the Web site \texttt{www.webmd.com} will be necessarily classified into an interest category related to health. Location-based and placement ads can evidently be mapped to any of the $n$ categories assumed in this work.

As commented in Sec.~\ref{sec:Background:AdServing}, the ad-serving process takes into account a wide range of variables when displaying an ad to a user on a given ad space.
These variables include tracking and profiling data about the user in question, the publisher being visited, the advertisers and their corresponding campaigns, and, depending on the ad-platform type, the bids of the ad-auction participants or the criteria of the ad platform itself to maximize its revenue.

\begin{figure*}[htb!]
\centering
\includegraphics[scale=0.65]{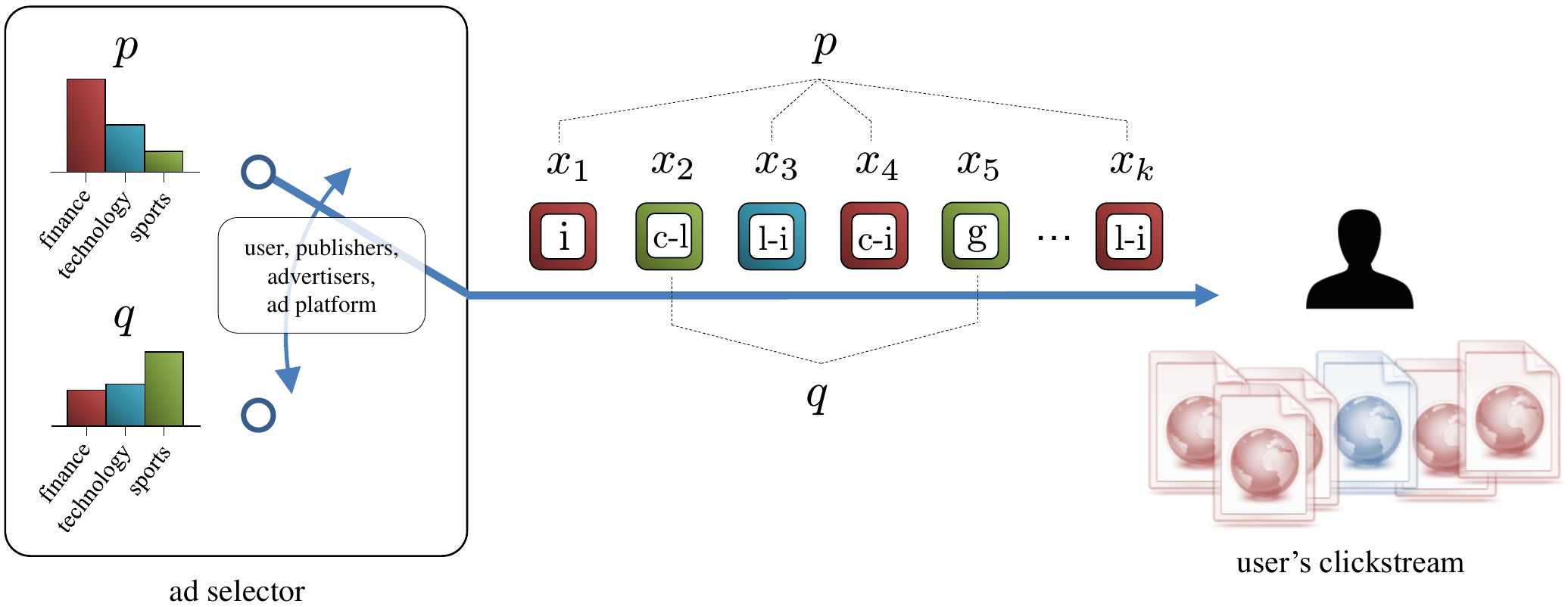}
\caption{An ad selector (e.g., a traditional ad platform) displays $k$ ads on the user's browser when navigating the Web. The interest categories of the delivered ads are modeled as a sequence of independent r.v.'s taking on values on $n=3$ categories. The observed categories, i.e., $(x_i)_{i=1}^k$, can be seen as generated by a source that commutes between the PMFs $p$ and $q$. The switching between interest-based ads (i.e., $\textnormal{``i''}$, $\textnormal{``c-i''}$, $\textnormal{``l-i''}$ and $\textnormal{``c-l-i''}$) on the one hand, and non-interest-based ads (i.e., $\textnormal{``c''}$, $\textnormal{``l''}$, $\textnormal{``g''}$ and $\textnormal{``c-l''}$) on the other, is determined by a number of parameters related to the user, publishers, advertisers and ad platform.}
\label{fig:ProbabilisticModel}
\end{figure*}

In our mathematical model, we characterize the ad-serving process conducted by an ad platform as a black box, whose inputs are the variables mentioned above, and whose outputs are the selected ads. We explained in the background section that traditional ad platforms are the ones selecting the ad to be displayed, while in RTB-based advertising the choice is made by the winning bidder, being an advertising agency or a traditional ad platform.
For the sake of conciseness and to avoid specifying the ad-platform model in each case, we shall henceforth use the term \emph{ad selector} to refer generically to the particular entity imposing the selection of an ad.

For each user and for each ad space, the outputted ads can be classified as content-, location-, and interested-based and generic,
according to the corresponding advertisers' targeting objectives. We note that, from these four classes of ads, we may only have eight possible combinations of those classes. Denoting each of the ad-classes by its first letter, the set of all such combinations is
$$\mathscr{G} = \{\textnormal{c}, \textnormal{l}, \textnormal{i}, \textnormal{g}, \textnormal{c-l}, \textnormal{c-i}, \textnormal{l-i}, \textnormal{c-l-i}\},$$
where the element $\textnormal{``c-l''}$ represents an ad that has been targeted based on content and location.
In other words, $\mathscr{G}$ includes all the combinations of targeting objectives an advertiser may choose.

We mentioned in Sec.~\ref{sec:Background:AdServing} that user profiles are essentially built from clickstreams, i.e., from the Web pages tracked.
For $k>>1$, let $(X_i)_{i=1}^k$ be the sequence of ads that an ad selector (e.g., a traditional ad platform) delivers to a particular user during several browsing sessions. Our characterization of this ad-delivery process stems from the intuitive observation that, if we were able to rule out all but the interest-based ads of such sequence, the empirical distribution~\cite{Cover06B} of the interest categories observed would naturally resemble, to a large extent, the user's browsing interests, or equivalently, their clickstream.

According to this observation and without loss of generality, we model the sequence of outgoing ads, classified into interest categories, as the output of an ad-source that \emph{alternates} between two probability distributions, namely

\begin{itemize}
  \item an interest-category distribution $p$ that reflects the knowledge the ad selector has about the \emph{user's interests};
  \item and another interest-category distribution $q$ that represents the complement of the former distribution and thus corresponds to (the interest categories of) those ads classified as non-interest-based, that is, \emph{contextual}, \emph{location-based} and \emph{generic}.
\end{itemize}

Naturally, the model described above captures only one aspect of the ad-serving process: it reflects the selection of the ads interest-categories within the set $\mathscr{X}$, a step that we model through the distributions $p$ and $q$ when the ad-class is respectively interest-based and non-interest-based. The proposed model is supported by the reasonable assumption that the accumulated interest categories of the interest-based ads will very likely approximate to the user's interests, or more precisely, the clickstream possessed by the ad selector.

Our model does not, therefore, capture other aspects of the ad-serving process like how a particular ad-class combination is chosen from $\mathscr{G}$. With it, however, we reflect the simple fact that the interest categories of the outgoing ads may be distributed according to either partial (or complete) user browsing data, or any other information which does not include those browsing data.
This simplified ad-serving model based on interest categories will allow us in the next subsection to estimate the ad-class chosen by the ad selector, or more accurately, whether the delivered ads are classified as interest-based or not. Fig.~\ref{fig:ProbabilisticModel} illustrates how we model this aspect of the ad-serving process.

\subsubsection{Binary Hypothesis Testing}
\label{sec:System:DetectionIB:Hypothesis}
\noindent
Assuming such model on the ad-platform's side, on the user's side we aim to determine if an ad, previously classified into an interest category, has been shown to the user based on their past Web-browsing interests or not. Formally, we may consider this as a binary \emph{hypothesis testing} problem~\cite{Cover06B} between two hypothesis, namely whether the data (i.e., the category of the displayed ad) has been drawn according to the distribution $p$ or $q$. Next, we elaborate on these two distributions.
Further details about the practical estimation of both PMFs are set forth in Sec.~\ref{sec:Implementation}.

\begin{figure*}[tb!]%
\centering
\includegraphics[scale=0.65]{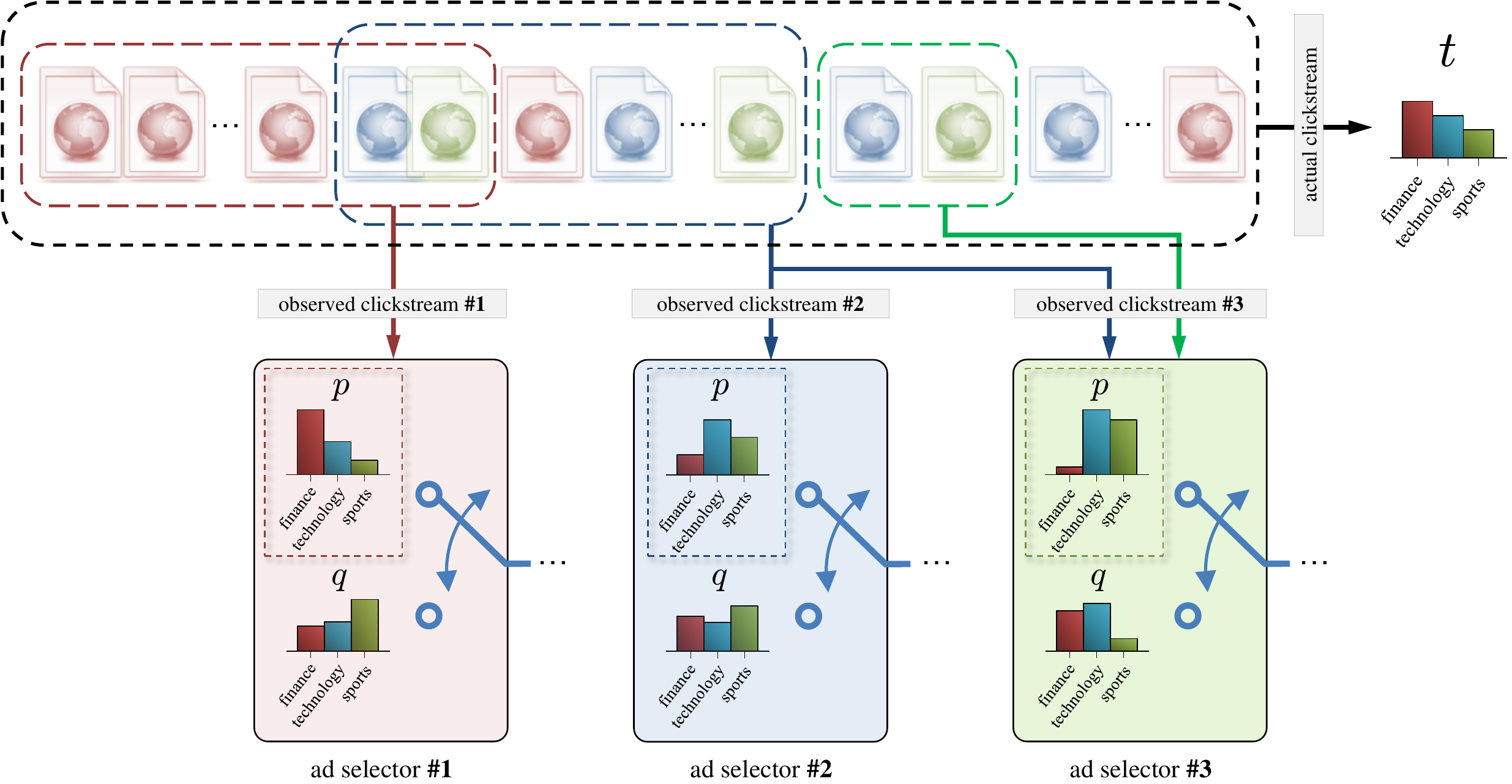}
\caption{We show how three ad selectors track a user through different Web sites.
The ad selectors 1 and 2 could represent two ad platforms overlapping their observed clickstreams.
This would reflect a common situation for large ad platforms like Google AdSense and OpenX.
The ad selector 3, on the other hand, could exemplify a small advertising company.
Because of its limited ability to track users on its own, this latter ad selector might decide to acquire tracking data from the ad selector 2.
Regardless of the data exchanged, however, none of the three ad selectors will be able to get the actual clickstream.}
\label{fig:ObservedActualClickStream}
\end{figure*}

Recall that, for a particular user and ad space, the \emph{ad selector} is the entity that ultimately decides which ad is shown to that user in that ad space. In the case of traditional ad platforms, the ad selector is the ad platform itself. In RTB, on the contrary, the ad selector is the bidder that wins the auction for displaying its ad, being an agency representing advertisers, a DSPs or a traditional ad platform.

As described in Sec.~\ref{sec:System:DetectionIB:Model}, the PMF $p$ represents the knowledge that such ad selector has about the user's browsing interests.
Henceforth, we shall refer to this distribution as the user's \emph{interest profile},
bearing in mind that it is specific to the ad selector in question.

In practice, these profiles are typically built from the tracked Web sites or \emph{observed clickstream}~\cite{Clickstream,Toubiana10SNDSS,CiscoGuide,LiuHotNet13,YanWWW09,AlyWWW12,PandeyCIKM11, Smith14B}.
The clickstream available to an ad selector, however, need not necessarily be the result of a direct tracking on the user.
For example,
ad platforms may track users on their own through their cookies;
and not satisfied with that, they may also wish to build upon tracking data from other ad platforms or trackers.
For the time being, we shall not specify how, in our model, the ad selector profiles a user from their clickstream. We shall only assume that profiles are represented as PMFs, as many works in the literature essentially do~\cite{Toubiana10SNDSS,Puglisi15DPM,LiuHotNet13,YanWWW09,AlyWWW12}.

Clearly, depending on the ability of the ad selector to track users throughout the Web (on its own or not),
the profile $p$ will resemble, to a greater or lesser extent, their actual interests. We denote by $t$ the interest profile resulting from the \emph{actual clickstream}, that is, all the Web sites visited by a user. We shall occasionally refer to $p$ and $t$ as the \emph{observed} and \emph{actual profiles}, respectively.
Fig.~\ref{fig:ObservedActualClickStream} extends the ad-targeting model depicted in Fig.~\ref{fig:ProbabilisticModel}, to reflect the fact that $p$ is constructed from the observed clickstream and thus may not capture the user's actual interest profile~$t$.

The distinction between these two profiles will also be employed later in Sec.~\ref{sec:Implementation} to reflect two possible scenarios regarding tracking and sharing of clickstream data: on the one hand, a paranoid scenario where users are tracked on every page they visit and such tracking data is exchanged among all entities serving ads.
And on the other hand, a baseline scenario where $p$ is fundamentally built from the clickstream an ad selector may get on its own, through cookies or other tracking technologies, without relying on tracking data from other sources.

In order to conduct our hypothesis testing, we shall also need to estimate the distribution $q$.
To this end, we consider an environment where no tracking is performed, similarly to when users enable the Web-browser's private mode.
Recall that this PMF is the interest-category distribution of those ads which are not profile-based, that is, those classified as $\textnormal{``c''}$, $\textnormal{``l''}$, $\textnormal{``g''}$ and $\textnormal{``c-l''}$.
Because, except for ad-placement, these ads will depend on the user's location and the pages visited during this free-tracking session,
$q$ will be specific to each particular user. To estimate this distribution on the user side, we shall capture the category of all ads received,
under the reasonable assumption that,
when users browse in private mode,
no browsing-interest data are leveraged to target the ads. In Sec.~\ref{sec:Implementation:Architecture:Components:Profiling}, we shall describe more specifically how this PMF will be estimated by our detector.

\subsubsection{Short-Term and Long-Term Interest Profiles}
\label{sec:System:DetectionIB:ShortLongProfiles}
\noindent
In previous sections, we pointed out that user interest profiles are mainly built from the categorization of the visited Web sites. We also commented that profiles are modeled essentially as PMFs, that is, as histograms of relative
frequencies of those visited sites across a set of interest categories. In this subsection, we briefly examine a crucial aspect of such user modeling, namely, we explore the importance that ad selectors may place on recent interests compared to those accumulated over a long time period.

From the perspective of profile-based targeting, the need to weight clickstreams is evident. A short recent history may be enough to direct products which do not require much thought, like buying a movie at Google Play. But other kind of transactions such as enrolling for an online university degree may need a longer browsing history to ensure a certain probability of conversion\footnote{In online marketing terminology, conversion usually means the act of converting Web site visitors into paying customers.}~\cite{PandeyCIKM11}.

Depending on the time window chosen, user profiles can be classified as \emph{short-term} and \emph{long-term} profiles. The former represent the user's current and immediate interests, whereas the latter capture interests which are not subject to frequent changes~\cite{Gauch07B}. In general, different interest-based marketing systems may contemplate different time windows for building profiles. Many commercial systems opt for relatively long-term profiles, while others capitalize on short, recent clickstreams. Some recent studies do not seem to agree on that, either. For example, \cite{PandeyCIKM11}~provides evidence that long browsing histories may lead to better targeting of users, while others show the opposite~\cite{YanWWW09}.

As we shall see in Sec.~\ref{sec:System:DetectionIB:Optimal:Robust}, our detection system will capture the uncertainty associated with the time window used by an ad selector. Since in practice it is impossible to ascertain this parameter, we shall consider uncertainty classes of user profiles. These classes will enable us to characterize the distinct options an ad selector might have chosen to create a profile, and will lead us to the design of an optimal robust detector.

\subsubsection{Optimal Detection of Interest-based Ads under Uncertainty}
\label{sec:System:DetectionIB:Optimal}
\noindent
In this section, we formulate the problem of designing an interest-based ad detector as a robust minimax optimization problem.
To this end, we essentially follow the methodology developed by~\cite{Boyd04B,Levy08B}.

Let $X$ be an r.v. modeling the category an ad belongs to.
Denote by $H$ the r.v. representing the two possible hypothesis about the distribution of the observed category $X$.
Let $H=1$ indicate that the ad is profile-based (first hypothesis), and $H=2$ it is not profile-based (second hypothesis).
Said otherwise, $X$ conditioned on $H$ has PMF $p$ when $H=1$ and $q$ when $H=2$.
For the sake of compactness, we denote by $P \in \mathbb{R}^{n \times 2}$ the matrix that has $p$ and $q$ as columns.

A \emph{randomized estimator} or \emph{detector} $\hat{H}$ of $H$ is a probabilistic decision rule determined by the conditional probability of $\hat{H}$ given $X$,
that is, $p_{\hat{H}|X}$.
The interpretation of such estimator is as follows: if $X$ is observed to have value $j$, the detector concludes $H=1$ with probability $p_{\hat{H}|X}(1|j)$, and $H=2$ with the complement of that probability.

A randomized detector also admits an interpretation in matrix terms,
in particular as an $\mathbb{R}^{2 \times n}$ matrix,
where the $j$-th column corresponds to the probability distribution of $\hat{H}$ when we receive an ad belonging to the interest category $j$.
Throughout this section, we shall conveniently use this matrix notation for estimators,
and denote by $D$ the matrix defining them.

The performance of a decision rule is usually characterized in terms of its detection and error probabilities.
We may capture this performance compactly by means of the matrix $M =D P$,
whose element $M_{ij}$ gives us the probability of deciding $\hat{H}=i$ when in fact $H=j$, that is, $p_{\hat{H}|H}(i|j)$.
The diagonal elements of this $2 \times 2$ matrix are the probabilities of correct guess.
The error probabilities are represented by the off-diagonal elements $M_{21}$ and $M_{12}$,
which yield the probabilities of a false negative and a false positive, respectively.
In our context, the former is the probability of concluding that the ad is not profile-based when actually it is;
and the latter is the probability of deciding the ad is interest-based when it is not.

Our aim is to design the matrix $D$ that defines the interest-based ads detector, so that certain performance criteria are satisfied.
Among other requirements, we might be interested in minimizing (maximizing) one of the error (detection) probabilities,
with a constraint on the complement of the objective probability.
Also, we could consider minimizing both error probabilities or a convex combination of them,
if some prior information about $p_H$ was available.

\paragraph{Robust Estimation}
\label{sec:System:DetectionIB:Optimal:Robust}
\noindent
Regardless of the criteria chosen, the problem of this design is that it requires the complete knowledge of the probability distributions defined by $P$.
As explained in the previous section, we may compute a reliable estimate of $q$ locally (i.e., on the user side),
but we cannot know how ad selectors construct the profile $p$ from their observed clickstream.
Some ad selectors may wish to target users based on their short-term interests,
some may rely on longer and relatively stable profiles to this end, and others may opt for both kind of models.
In any case, the time window/s employed by an ad selector is what determines the profile/s that will be used for ad targeting.
Because this information is unknown, having a precise specification of the distribution $p$, or estimating it reliably, is therefore infeasible.

\begin{figure}[tb!]%
\centering
\includegraphics[width=\columnwidth]{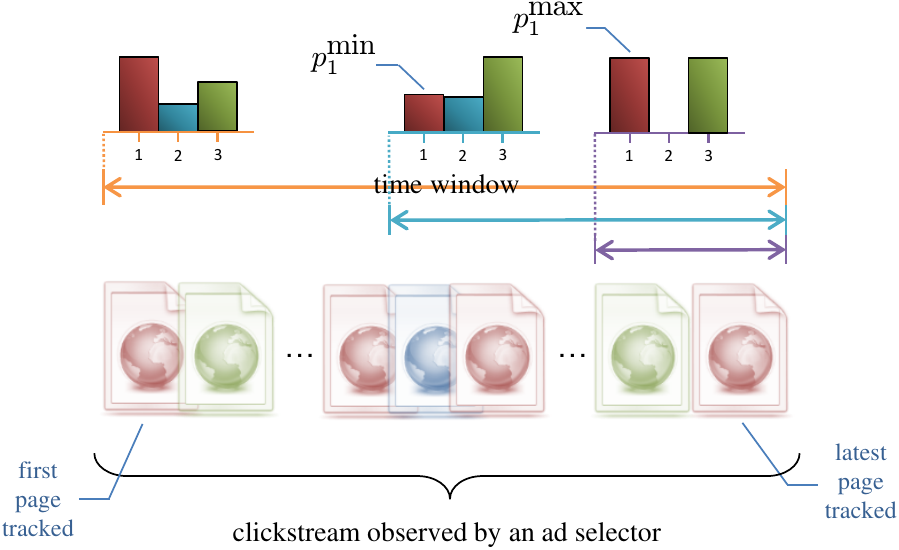}
\caption{Ad selectors may create interest profiles based on the Web pages tracked.
Our detector captures all possible options an ad selector may consider to compute those profiles from the tracked pages.
All these options are directly related to the time window/s chosen, or equivalently, the number of pages taken from the observed clickstream.
We model these possible choices as intervals between minimum and maximum interest values per category.}
\label{fig:Uncertainty}
\end{figure}

The problem of estimating a distribution under uncertainty has also been encountered in other fields and applications such as
signal processing~\cite{Zoubir12SP}, portfolio optimization~\cite{Nguyen09PhD} and communications networks~\cite{Yang08INFOCOM}.
In all these cases, the probability distributions are frequently specified to belong to sets of distributions, typically called \emph{uncertainty classes}.
In our case, the uncertainty class of $p$ is given by the minimum and the maximum lengths of the time windows an ad selector may define
to model short-term and long-term interests.
In practice, the maximum length might correspond to the entire clickstream,
whereas a minimum reasonable time window for short-term profiles might be one day~\cite{PandeyCIKM11, AlyWWW12}.

For $i=1,\ldots,n$, we denote by $p_i^{\textnormal{max}}$ the maximum interest value $p_i$ estimated by the ad selector,
over \emph{all} possible time windows ranging from one day to the whole observed clickstream\footnote{In Sec.~\ref{sec:Implementation:Architecture:Components:Profiling}, we shall see that a maximum time window of 1.5 months may be sufficient.}.
We define $p_i^{\textnormal{min}}$ analogously,
and intuitively model the uncertainty about the distribution $p$ as intervals between these upper and lower bounds.
More specifically, we define the set of possible interest profiles as
\begin{equation}\label{eq:unclass}
\mathcal{P} = \{p : p^{\textnormal{min}} \preceq p \preceq p^{\textnormal{max}}, \oone^{\oT}p= 1, p \succeq 0 \},
\end{equation}
where the symbol ``$\preceq$'' indicates componentwise inequality,
and the last inequality and the equality reflect the fact that $p$ must be a PMF.

At a conceptual level, the polyhedron $\mathcal{P}$ captures all the possible profiles that an ad selector may have built by adding incremental
observations of one Web site to the interests model.
By computing the maximum and minimum observed interests over all these incremental models, and
by defining intervals of interest values between these two extremes,
we obtain an uncertainty class that reflects \emph{any} possible decision made by the ad selector regarding the time window.
We would also like to stress that the uncertainty class $\mathcal{P}$ likewise includes
the possibility that an ad selector may be using more than one profile ---with different time windows--- for a same user.
Fig.~\ref{fig:Uncertainty} illustrates the uncertainty around the selected time window/s.

One possible way to devise an estimator when a probability distribution is specified to belong to an uncertainty class is to
contemplate the \emph{worst-case} performance over this class.
The resulting decision rule is then said to be \emph{robust} to the uncertainties in the probability distribution~\cite{Kouvelis96}.
Following the notation of~\cite{Boyd04B}, we define the worst-case performance matrix $M^w$ associated with a robust detector as
$$M_{ij}^w = \sup_{p\in\mathcal{P}} M_{ij},$$ for $i,j=1,2$, with $i\neq j,$
and $$M_{ii}^w = \inf_{p\in\mathcal{P}} M_{ii},$$ for $i=1,2$.
In general terms, the off-diagonal elements of this matrix give us the largest probability of errors over all $p\in\mathcal{P}$.
The diagonal entries, on the other hand, yield the smallest possible detection probabilities.
Based on the latter probabilities, we may define the \emph{worst-case error probability} as $P_i^w = 1 - M_{ii}^w$,
which represents the largest probability of error over the uncertainty class when $H=i$.
Clearly, we note that $M_{12}^w= M_{12}$ and $M_{22}^w=M_{22}$, as in our case the uncertainly is just in~$p$.

\paragraph{Minimax Design}
\label{sec:System:DetectionIB:Optimal:Minimax}
\noindent
In this subsection, we specify the design of a robust interest-based ad detector,
and formulate the hypothesis test problem between $H_1$ and $H_2$ as a linear program (LP).

Based on the error and detection probabilities shown in the previous subsection, various designs can be developed.
Some classical optimality criteria are the Bayes, Neyman-Pearson and minimax designs~\cite{Levy08B}.
In this work, we consider a robust minimax approach that minimizes the worst-case error probability, over the two hypotheses.
We adopt this approach because, in our attempt to detect interest-based ads, both error probabilities are equally important.

According to this design criterion, the proposed robust minimax detector is given by the matrix $D$ that solves the optimization problem
\begin{equation}
\label{eq:Minimax}
\min \max_{i=1,2} P_i^w.
\end{equation}
Let $\tilde{d}^{\,\oT}$ be the first row of $D$, that is, the conditional probabilities $p_{\hat{H}|X}(1|j)$ for $j=1,\ldots,n$.
We show in the Appendix~\ref{Appendix:LP}
that~\eqref{eq:Minimax} is equivalent to the following optimization problem
in the variables $\lambda,\mu, \tilde{d}\in \mathbb{R}^n$ and $\nu\in \mathbb{R}$:
\begin{equation}\label{eq:MinimaxLP}
\begin{aligned}\quad
\textnormal{maximize}     &\quad\quad \zeta \\
\textnormal{subject to}   &\quad\quad \mu^{\oT}p^{\textnormal{min}} - \lambda^{\oT} p^{\textnormal{max}} + \nu \geqslant \zeta,\\
                                            &\quad\quad 1 - \tilde{d}^{\,\oT}q  \geqslant \zeta, \\
                                            &\quad\quad \mu - \lambda + \nu\oone \preceq \tilde{d},\\
                                            &\quad\quad \lambda \succeq 0,   \mu \succeq 0,\\
                                            &\quad\quad 0\preceq\tilde{d}\preceq\oone.
\end{aligned}
\end{equation}
The strength of recasting~\eqref{eq:Minimax} as an LP lies in that it allows us to resort to extremely efficient and powerful methods to compute the optimal detector.
This is of a great practical relevance as we aim to provide such interest-based detection functionality on the user side, as a stand-alone software operating in real-time, i.e., while the user browses the Web.
Sec.~\ref{sec:Implementation} will give further details about the optimization library used for this computation.
The feasibility of this optimization problem is shown in Appendix~\ref{Appendix:Feasibility}.

\subsection{Detection of Profile Uniqueness}
\label{sec:System:DetectionUnique}
\noindent
In the previous subsection, we provided the design of a robust interest-based detector whereby users may learn to what extent their browsing profiles are exploited to serve them ads.
This subsection investigates another crucial aspect related to behavioral targeting, namely, if the profiles collected by the advertising companies might reveal \emph{unique} browsing patterns.

The importance of this aspect lies in the potential risk of reidentification from unique, non-personally identifiable data,
as illustrated, for example, by the \emph{AOL} search data scandal~\cite{AOL06}\footnote{\emph{AOL} user No. 4417749 found this out the hard way in 2006, when \emph{AOL} released a text file intended for research purposes containing twenty million search
keywords including hers.
Reporters were able to narrow down the 62-year-old widow in Lilburn, Ga., by examining the content of her search queries~\cite{AOL06}.}.
In our context, the risk of profiling goes hand in hand with the risk of reidentification,
especially when considered in the context of additional information obtainable from a user such as their location, accurate navigation timing and aspects related to the Web browser and operating system.
When the profile is added also to the wealth of data shared across numerous information services, which a privacy attacker could observe and cross-reference, such attacker might eventually find out, even if in a statistical sense, the user's real identity.

Having motivated the risk of profile uniqueness, this subsection describes how to detect if the ads delivered to a user may have been generated as a result of a common browsing pattern, or conversely, to a browsing history that deviates from a typical behavior.
To this end, we first provide a brief justification of KL divergence as a measure of the uniqueness of a profile, or equivalently, its commonality.
The rationale behind the use of divergence to capture this aspect of a profile
is documented in greater detail in~\cite{Parra13FGCS,Rebollo11SECTECH}.
Afterwards, we examine how to estimate this information-theoretic quantity.

Although we mentioned in Sec.~\ref{sec:System:Preliminaries} that the KL divergence is not a proper metric, its sense of discrepancy between distributions allows an intuitive justification as a measure of profile commonality. Particularly,
whenever the profile observed by an ad selector diverges too much from the average profile of all tracked users, the ad selector will be able to ascertain whether the interests of the user in question are atypical, in contrast to the statistics of the general population.

A richer justification arises from Jaynes' celebrated \emph{rationale on entropy maximization methods}~\cite{Jaynes82P,Jaynes57PRS2},
which builds on the method of types~\cite[\S 11]{Cover06B}, a powerful technique in large deviation theory.
Leveraging on this rationale,
the relative entropy between an observed profile and the population's profile may be considered as a measure of the uniqueness of the former distribution within such population.
The leading idea is that the method of types establishes an approximate monotonic relationship
between the \emph{likelihood} of a PMF in a stochastic system and its divergence with respect to a reference distribution,
say the population's.
Loosely speaking and in our context, the lower the divergence of a profile with respect to the average profile, the more likely it is, and the more users behave according to it.
Under this interpretation, the KL divergence is therefore interpreted as an (inverse) indicator of the commonness of similar profiles in said population.

Having argued for the use of KL divergence as a measure of profile commonality,
next we elaborate on the uncertainty to estimate this divergence value.
Recall from Sec.~\ref{sec:System:DetectionIB:ShortLongProfiles} that ad selectors may construct profiles in multiple ways from the observed clickstream.
Just as we did with the design of the interest-based ad estimator,
we proceed by considering a worst-case uniqueness estimate on the space of possible profiles built by an ad selector.
Denote by $\bar{p}$ the \emph{population's interest profile}.
Formally, for each user and ad selector, we define the \emph{minimum uniqueness} over all such profiles as
\begin{equation}\label{eq:uniqueness}
u_{\textnormal{min}} =\inf_{p\in\mathscr{P}} \oD(p\,\|\,\bar{p}),
\end{equation}
which gives a measure of profile commonness that allows for the uncertainty inherent in the time window used by an ad selector.

The divergence-minimization problem above captures a worst-case scenario regarding profile commonality.
In particular, it
tells us how peculiar our interests might be, as seen by an ad selector.
For any ad selector, the value $u_{\textnormal{min}}$ (on the interval $[0,\infty)$ bits) will clearly vary over time as the user browses the Web.
From the point of view of comprehensiveness,
however, the information conveyed each time by this \emph{absolute} uniqueness value may not be informative enough to the user.

To help the user interpret a given $u_{\textnormal{min}}$ value, we consider making it \emph{relative} to a population of users.
In doing so, users can compare their profile uniqueness values with those of other users of our Web-browser extension,
and thus gain a broader perspective of how they are profiled.
Also, users may utilize this information to define consequent ad-blocking policies.
Later in Sec.~\ref{sec:Implementation}, we shall describe the exchange of information between users of our system and a central repository to estimate those relative profile-uniqueness values.

\section{``MyAdChoices'' --- an Ad Transparency and Blocking Tool}
\label{sec:Implementation}
\noindent
This section describes \emph{MyAdChoices}, a prototype system that aims to bring transparency into the ad-delivery process, so that users can make an informed and equitable decision regarding ad blocking.
The proposed system provides two main functionalities.
Enabled by the interest-based ad detector and the profile-uniqueness estimator designed in Sec.~\ref{sec:System},
the ad-transparency functionality allows users to understand what is happening behind the scenes with their Web-browsing data.
The ad-blocking functionality, on the other hand, permits users to react accordingly, in a flexible and non-radical manner.
This is unlike current ad-blocking technologies, which simply block or allow all ads.
\emph{MyAdChoices} does not only consider these two extremes,
but the interesting and necessary continuum in between.
With this latter functionality, users can indicate the type of ads they wish to receive or, said otherwise, those which they want to block.
By combining both functionalities and thus providing transparency and fine-grained control over online advertising,
the proposed system may help preserve
the Internet's dominant economy model,
currently threatened by the rise of simple, radical ad blockers.

This section is organized as follows. Sec.~\ref{sec:Implementation:Functionalities} first elaborates on the
ad transparency and blocking functionalities provided by our system.
Afterwards, Sec.~\ref{sec:Implementation:Architecture} describes the components of a system architecture that implements these two functionalities.

\subsection{Main Functionalities}
\label{sec:Implementation:Functionalities}
\noindent
Our system brings \emph{transparency} to two central aspects of behavioral ad-serving.
On the one hand, it allows users to know if the information gathered about their browsing interests may have been utilized by the advertising industry to target them ads.
Specifically, our system lets the user know if the received ads may have been generated according to their browsing interests or, more accurately, to the profiles that ad selectors may have about them.
On the other hand, it provides insight into the browsing profiles that ad selectors may have inferred from the pages tracked.
In particular, \emph{MyAdChoices} shows a worst-case, profile-uniqueness value for each ad selector, and the interest category of the ads received.

With regard to the \emph{ad-blocking} service,
our system contemplates the following user-configurable parameters:

\begin{itemize}
  \item \textbf{Ad interest-category}.
We offer users the possibility to filter ads by interest category.
For example, a user could block ads belonging to certain sensitive categories like pornography and health.

  \item \textbf{Ad class}.
This parameter enables users to block either the interest-based ads or the non-interest-based
ads, for all ad interest-categories or for a subset of them.

  \item \textbf{Profile uniqueness}.
Users may decide blocking the ads delivered by those ad selectors that may have compiled very unique, and thus potentially re-identifiable, profiles of their browsing habits.

  \item \textbf{Retargeting}. Last but not least, users can decide to block retargeted ads, that is, ads coming from advertisers that have been previously visited by the user (see Sec.~\ref{sec:Background:AdClasses}).

\end{itemize}

\subsubsection{Examples of Ad-Blocking Policies}
\label{sec:Implementation:Functionalities:Examples}
\noindent
This subsection provides a couple of simple but insightful ad-control policies that aim to illustrate the parameters described in the previous subsection.
These examples are prefaced by a general definition of ad-filtering policy, inspired from the field of access control.

\begin{definition}[Ad-blocking policy]
\label{def:policy}
A \emph{policy} $\ensuremath{\mathit{pol}}$ is a pair $(\ensuremath{\mathit{AC}},\ensuremath{\mathit{sign}})$, where
$\ensuremath{\mathit{AC}}$ is an {\em ad constraint}, and $\ensuremath{\mathit{sign}}\in\{+,-\}$ models an \emph{action} to be taking when an ad meets that constraint.
An ad constraint is represented by a triple $(I, i, u)$, where $I\in\{0,1\}$ indicates if an ad is interest-based or not, $i\in\mathscr{X}$ is an interest category, and $u_{\textnormal{min}}$ denotes a requirement of minimum profile uniqueness.
\end{definition}

An ad constraint represents the set of ads belonging to an interest category $i$, which are classified as interested-based (or not), and which have been delivered according to a profile with minimum uniqueness given by $u_{\textnormal{min}}$.
On the other hand, $\ensuremath{\mathit{sign}}$ denotes if the ad must be blocked (-), or displayed on the user's browser (+).

Because the support for positive and negative policies may cause conflicts (i.e., we may have an ad constraint satisfying both positive and negative policies),
a conflict-resolution mechanism must be enforced.
The literature of access control provides several approaches to tackle such conflicts.
A comprehensive survey on this topic is~\cite{Ferrari00B}.
Here, for simplicity, we assume that negative policies prevail,
since this approach provides stronger guarantees with regard to the risk of displaying unappropriate ads.
Other conflict resolution policies, however, could also be readily integrated.

Two examples of policies are given next. In these examples, we refer to some of the interest categories used by the proposed system (see Sec.~\ref{sec:Implementation} for more details). For brevity, in this section we shall denote the relevant categories by its name.
Also, for simplicity and clarity, in the examples we shall keep using the policy formal notation introduced in Definition \ref{def:policy}.
We note, however, that this notation, describing how policies are actually implemented in the system, must be made transparent in the front end both to improve usability and to help users specify policies reflecting as much as possible their preferences.
As we shall explained in Sec.~\ref{sec:Implementation:Architecture:Components:Policy}, several strategies will be devised for this purpose, e.g., the use of textual labels instead of numeric values.

\begin{example}[Policies for allowing certain personalized ads]
\label{example:1}
Alice had planned to visit New York City (NYC) for her holidays.
Some days ago she bought her flight tickets and booked her hotel, all through the Internet.
During the following days, she visited several Web sites in search of sightseeing tours and day trips.
As she browsed the Web, the ads displayed in her browser became increasingly related to her upcoming trip.
Alice is now fed up with ads on hotels in NYC, so she is considering installing AdBlock Plus to block them all. However, she appreciates the value and usefulness of behavioral targeting, and because she has not decided her itinerary yet, she still wants to receive personalized ads associated with the categories 1 (``Travel\textbackslash Trains") and 2 (``Travel\textbackslash Theme parks").
Consequently, Alice specifies the following policies:
\end{example}
\begin{itemize}
\item $\ensuremath{\mathit{pol}}_{1} = ((c_1, 1,\cdot),+)$,
\item $\ensuremath{\mathit{pol}}_{2} = ((c_2, 1,\cdot),+)$,
\end{itemize}
where the symbol ``$\cdot$'' means that the value of the parameter in question is not specified.

\begin{example}[Policy for balancing personalization and privacy]
\label{example:2}
Bob works in a dietetics and nutrition shop. As part of his work, he sometimes consults pages about health and fitness. Occasionally, and when nobody sees him, he spends some time checking Web sites related to his recently diagnosed fibromyalgia's disease. Some days ago he was shocked when a couple of ads on biological treatments for his disease popped up while he was browsing the Web. Since then Bob is very concerned that related ads may be displayed when his workmates look over his monitor. However, despite his worries, he does not wish to resort to the typical ad-blocking
plug-ins, as such personalized-ads services also helps him keep abreast of the newest products and trends in his work. To strike a balance between privacy and personalization, Bob specifies a filter that blocks profile-based, health-related ads only when his browsing profile reflects relatively atypical interests. In particular, he defines the following policy:
\end{example}

\begin{itemize}
\item $\ensuremath{\mathit{pol}}_{1} = ((c_3, 1, \pi_{u_{\textnormal{min}}}\geqslant 25\%),-)$,
\end{itemize}
where the category 3 corresponds to ``health \& fitness'',
and $\pi_{u_{\textnormal{min}}}$ denotes the percentile value of $u_{\textnormal{min}}$.

\begin{figure*}[tp!]
\centering
\includegraphics[scale=0.90]{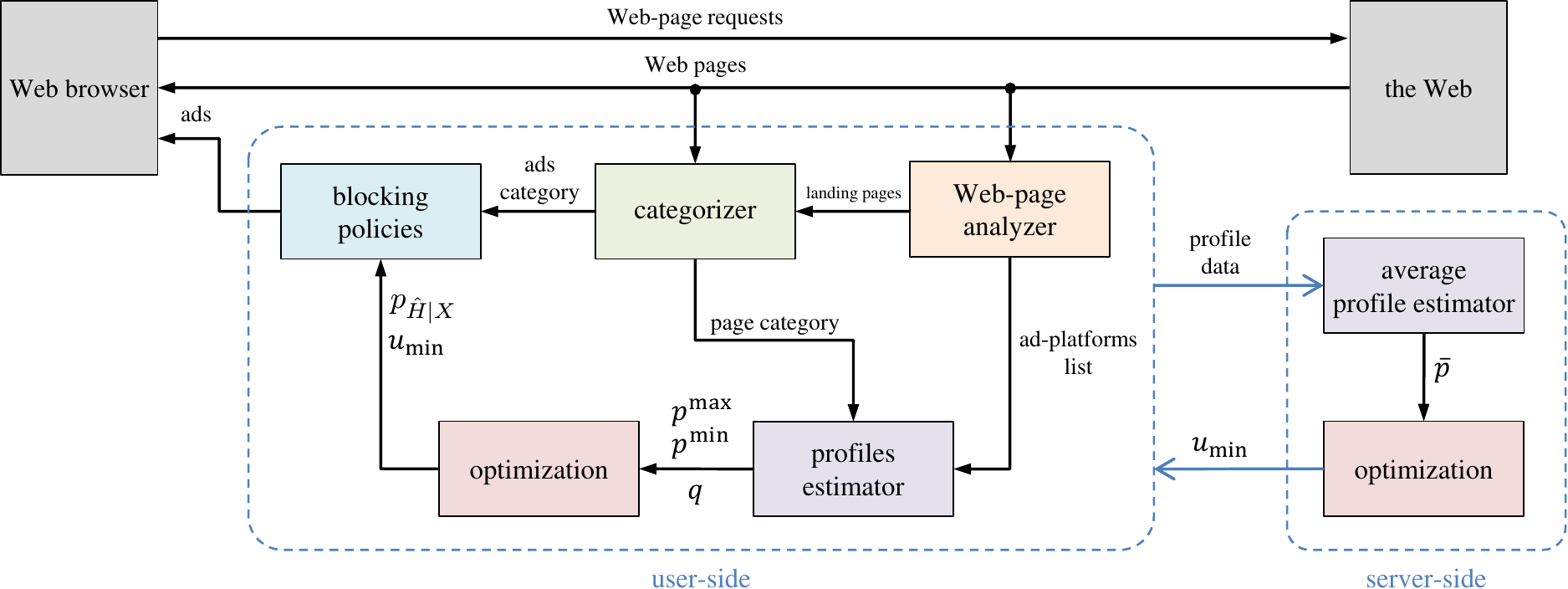}
\caption{Internal components of the proposed architecture.}
\label{fig:Architecture}
\end{figure*}

Lastly, we would like to emphasize the topicality and appropriateness of this latter example, with an extreme case in which a cancer patient reported numerous Facebook ads for funeral companies after having searched for his recently diagnosed disease~\cite{Woollaston15DM}.

\subsection{System Architecture and Implementation Details}
\label{sec:Implementation:Architecture}
\noindent
In this section, we describe the components of a system architecture that implements the two functionalities specified in Sec.~\ref{sec:System}.
The proposed system has been developed as a Web-browser extension and is available for Google Chrome\footnote{Currently,
the tool is in beta version and can be downloaded at \protect\url{https://myrealonlinechoices.inrialpes.fr} under request.}.
It is worth emphasizing that this extension not only provides transparency and ad-blocking services in real-time, but also operates as a stand-alone system, i.e.,
it performs all computations and operations locally, without the need of any infrastructure or external entity to this end.
The only exception is the computation of the minimum profile-uniqueness value, which is not done on the user side, as it requires the average profile of the population~$\bar{p}$. As we shall elaborate later on in Sec.~\ref{sec:Implementation:Architecture:Components:Optimization}, this particular service is provided only if the user accepts sharing profile data with the \emph{MyAdChoices} server.

\subsubsection{Assumptions}
\label{sec:Implementation:Architecture:Limitations}
\noindent
Before proceeding with the description of the system architecture, we examine the assumptions made in implementing the interest-based ad detector and the profile-uniqueness estimator designed in Secs.~\ref{sec:System:DetectionIB} and~\ref{sec:System:DetectionUnique}.

Our first assumption is related to the impossibility of finding out, with absolute certainty, the browsing information that ad selectors have about users. In Sec.~\ref{sec:System:DetectionIB} we called this information the observed clickstream, and defined it more precisely as the sequence of Web pages the ad selector knows that the user visited.
By observing the third-party network requests, our browser extension is able to capture the pages ad platforms may track through HTTP cookies or other more sophisticated methods like Web-browser fingerprinting.
Nevertheless, we cannot know if this is all the information available to them, i.e., if those pages account for their observed clickstreams or not
---ad selectors and Web trackers may also exchange their tracking data, for example, through cookie matching,
a practice that appears to be much more common than those direct tracking methods~\cite{Olejnik14NDSS,Olejnik15PhD,Acar14CCS}.
The fact that a cookie-matching protocol is executed between two entities does not imply, however, that they end up exchanging their tracking data.
There is an obvious incentive to aggregate information and gain further insight into a user's browsing history, but since
this exchange does not go through the user's browser,
we cannot safely conclude that it is made.

In the case of RTB, the bid requests sent by an ad platform may enable the auction participants to track a given user.
Since the winning bidder (i.e., the ad selector) is the one serving the ad, our system can easily flag the corresponding page as being tracked by this bidder.
The problem, however, is that we cannot ascertain if this ad selector could have received other bid requests for this user (while visiting other pages), and thus could have tracked them across those pages.
Ad platforms typically permit bidders to build profiles \emph{only} from the auctions they win,
but, actually, nothing precludes them technically from exploiting such tracking data.
In short, because there is no way of knowing the recipients of those requests and the use they make of such data, our knowledge of the sites tracked through RTB is limited to those sites where the ad selector serves an ad.

In this work, we address all such limitations by considering two scenarios in terms of tracking and sharing of clickstream data:

\begin{itemize}
  \item a \emph{baseline} scenario, where the system operates with the clickstream data that, according to our observations, the ad selector may have. That is, we assume that the observed clickstream of an ad selector matches the tracking data of which we are aware,
      and therefore we ignore any possible sharing of tracking information with other entities.
In practical terms, our Web-browser extension will compile this clickstream
by examining if the ad selector is present, as a third-party domain, on the pages visited by the user.
In other words, we shall assume that all third-party domains present on a page may track a user's visit to such page.
By doing so, we will be able to capture the sites where an ad selector has embedded a link (through the corresponding publishers),
and those pages where it has won the right to serve an ad through RTB.

  \item a \emph{paranoid} scenario in which we assume Web tracking is ubiquitous and clickstream information is shared among all entities participating in the ad-delivery process.
      In this case, we consider that the observed clickstream coincides with the actual clickstream, i.e., with the sequence of all pages a user has visited.
      We acknowledge, nevertheless, that there may not be ad companies and trackers on certain pages and thus a complete, accurate actual profile might not be captured in practice.
      \end{itemize}

We would like to underline that the two scenarios described above refer \emph{solely} to the user-tracking data available to ad selectors.
Put differently, our system does not consider any interests data and personal information that users could have \emph{explicitly} conveyed to these entities, and that could be utilized for ad-targeting purposes.

Having specified the two modes of operation of our system, next we introduce our second assumption,
which concerns the way in which ad selectors construct user profiles from the observed clickstreams.
In Sec.~\ref{sec:System:DetectionIB:Hypothesis} we assumed that ad selectors model profiles as PMFs, essentially in line with a great deal of the literature on the field.
To compute such distributions in practice, our system assumes, with a slight loss of generality,
that ad selectors employ maximum-likelihood (ML) estimation~\cite{Schervish95B}.
We would like to stress that this is, by far, the most popular method of parameter estimation in statistics.

Our third and last assumption has to do with the topic categorization of the Web content.
We shall consider that the categorizer used by our system coincides, to a large degree,
with the one employed by ad platforms\footnote{Ad platforms are the ones classifying the content of a page. In RTB advertising, they typically include the category of the publisher's page in the bid requests.}.
This implies that both our extension and ad platforms rely largely on the same predefined set of interest categories and
the same categorization algorithm, so that any page visited by the user is classified into the same category by both the
proposed system and the ad platforms tracking this visit.
We believe this is a plausible assumption since our categorization algorithm builds on the standard topic taxonomy
developed by the Interactive Advertising Bureau~\cite{IAB15Web},
an organization that accounts for the vast majority of online advertising companies in the US.

\subsubsection{Components}
\label{sec:Implementation:Architecture:Components}
\noindent
This section provides a functional description of the main components of our prototype system architecture, justifies the design criteria, and gives \emph{some} key, low-level implementation details.
Fig.~\ref{fig:Architecture} depicts the implemented architecture,
which consists of two main parts, the user side and the server side.
The latter is in charge of computing the values of minimum uniqueness per ad selector.
Because this requires obtaining $\bar{p}$,
said computation is carried out only if the user accepts sharing their profile data with our server.
The rest of functionalities and processing is conducted entirely on the user side.
We analyze the components of both sides in the following subsections.

\paragraph{Profiles Estimator}
\label{sec:Implementation:Architecture:Components:Profiling}
\noindent
On the user side, this module aims at estimating (1) the set $\mathcal{P}$ of possible user profiles an ad selector may have assigned to a user; and (2) the distribution $q$ of the interest categories of those ads classified as non-interest-based.
It is important to stress that, regardless of the scenario assumed (i.e., baseline or paranoid), the estimation of $q$ must be carried for \emph{each} ad selector.
In the former scenario, the computation of $p$ is also necessary per ad selector. However, since the latter scenario considers that the observed clickstreams of all ad selectors match the user's actual clickstream, we just assume that~$p=t$.

As explained in Sec.~\ref{sec:System:DetectionIB:Hypothesis}, the estimation of the PMF $q$ requires a browsing session where the user is not tracked.
Our current version of the plug-in implements this free-tracking session by means of the browser's private or incognito mode,
a browser's feature that, among other functionalities, prevents tracking through HTTP and Flash cookies.
We acknowledge, however, that ad selectors might also follow users' visits as a result of using super cookies, respawning~\cite{Soltani10IIPM,Kamkar10}, canvas fingerprinting~\cite{Mowery12W2SP} or simply their IP addresses.
Nevertheless, since these tracking mechanisms are either very infrequent or rather inaccurate,
we may reasonably assume that the browser's incognito mode closely matches an untracked session, if not completely.
In fact, recent studies indicate that the prevalence of these more sophisticated tracking methods is just 5\% on top Alexa 100\,000 sites~\cite{Acar14CCS}.
In short, we shall therefore consider that the PMF $q$ estimated this way effectively reflects the ad-topic distribution when the user is seen by the ad selector as a new user, and thus the ads can only be location-based, contextual and generic.

In practical terms, there is a difference between the estimation of $p$ and $q$.
In the latter case, it is conducted from the ads the browser receives during such incognito mode.
In the case of $p$, or equivalently $\mathcal{P}$, the estimation is carried out from the pages the ad selector is able to track,
on its own and/or through other sources of data.

One of the difficulties in estimating these two distributions is that, while $q$ requires browsing in such free-tracking session,
the PMF $p$ must reflect the pages tracked by any potential ad selector.
An approach to dealing with this incompatibility consists in alternating between the incognito and the normal modes on a regular basis.
The problem with such approach, however, is that users might be reluctant to browse in the private mode for the time needed to compute and update the PMFs $q$ of a sufficient number of ad selectors.

\begin{table*}[tb!]
\centering
\caption{Top-level interest categories.}
\label{tab:TopLevelCategories}
\footnotesize
  \begin{tabular}{cccc}
    \toprule
        adult                                & economics                       & hobbies \& interests                         &  politics \\
    agriculture                     &  education                                     &  home                                     & real estate \\
    animals                         & family \& parenting                              & law                              & religion \\
    architecture                  & fashion                                                  & military                                         & science\\
    arts \& entertainment  & folklore                                         & news                                 & society \\
    automotive                    & food \& drink                                 & personal finance                                                 & sports\\
    business                        & health \& fitness                     & pets                           & technology \& computing\\
    careers                          & history                               & philosophy                       & travel\\
    \bottomrule
  \end{tabular}
\end{table*}

\begin{table*}[tb!]
\caption{Subcategories corresponding to three top-level categories.}
\label{tab:SubLevelCategories}
\centering
\footnotesize
\begin{tabularx}{.8\textwidth}{ >{\bfseries }cX}
\toprule
\textbf{Top-level category} & \makebox[9.6cm][c]{\textbf{Bottom-level category}} \\ [0.8ex]
\toprule
arts \& entertainment  & animation, celebrities, comics, design, fine art, humor, literature, movies, music, opera, poetry, radio, television,  theatre and video games. \\\midrule
health \& fitness & alternative medicine, anatomy, asthma, autism, bowel incontinence, brain tumor, cancer, cardiac arrest, chronic pain, cold \& flu, deafness, dental care, dermatology, diabetes, dieting, epilepsy, exercise, eye care, first aid, heart disease, HIV/AIDS, medicine, men's health, mental depression, nutrition, orthopedics, pediatrics, physical therapy, psychology \& psychiatry, senior health, sexuality, sleeping disorders, smoking cessation, stress, substance abuse, thyroid disease, vitamins, weight loss and women's health. \\\midrule
personal finance & banking, credit, debt \& loans, cryptocurrencies, financial news, financial planning, insurance, investing, retirement planning, stocks and tax planning.\\
\bottomrule
\end{tabularx}
\end{table*}

Motivated by this, the user-side architecture simultaneously estimates both distributions
by revisiting, in the incognito mode and in an automated manner, a \emph{fraction} $\rho$ of the pages browsed by the user.
In practical terms, each revisit is made by opening a new minimized window in the private mode. We proceed this way because
we want to avoid the tracking
among different tabs in the same incognito mode.
We admit, nonetheless, that this approach might have a non-negligible impact on these two aspects:
first, in terms of the traffic overhead incurred;
and secondly,
it may penalize advertisers to some degree, since the ads received in the free-tracking session will obviously not be presented to the user.
Currently, the proposed system operates with a revisit ratio of $\rho= 25\%$.
Although this reduction in the number of revisits undoubtedly comes at the cost of inaccuracy in the estimation of $q$,
we believe that it may account for an acceptable overhead in terms of traffic overhead and advertising impact.
As a side note, we would like to stress that the impact of such revisits is, from a usability perspective, almost imperceptible.

After examining the Web-browsing conditions in which $p$ and $q$ are obtained,
next we describe more concrete aspects related to the estimator of these distributions.

As mentioned in Sec.~\ref{sec:Implementation:Architecture:Limitations}, this work assumes that ad selectors rely on ML estimation, a simple estimation method widely common in many fields of engineering.
Let $m$ denote the total amount of ads received (pages visited), and $m_i$ the number of those ads (pages) which belong to the
interest category $i$.
Recall that the ML estimate of a PMF is defined as
\begin{equation*}
q_i = \frac{m_i}{m},
\end{equation*}
for $i=1,\ldots,n$.

In order to make a decision on whether the displayed ads are interest-based or not, our ML estimator requires observing the same minimum number of pages $w_{\textnormal{min}}$ needed by an ad selector to model short-term interests. Several studies point out that the smallest time window that advertising companies might use for such modeling is one day (see Sec.~\ref{sec:System:DetectionIB:ShortLongProfiles}). According to these studies and to the average number of pages browsed by a user per day~\cite{Nielsen10TR}, we set $w_{\textnormal{min}}=87$. On the other extreme, in line with the works cited in that section, we consider that the largest clickstream used to model long-term interests is 8 weeks. We then set $w_{\textnormal{max}} = 3\,915$.
To estimate $q$, we proceed analogously, by establishing a sliding window of this same length.

Lastly, on the server side, our architecture aims at computing, for each user willing to share profile data with the server, the average profile and the uncertainty class of each ad selector.

\paragraph{Web-Page Analyzer}
\label{sec:Implementation:Architecture:Components:PageAnalyzer}
\noindent
This block aims at obtaining certain information
 about (1) the Web pages browsed by the user and (2) the ads displayed within those pages, \emph{both} in the tracked and in the incognito sessions. Specifically, when the browser downloads a page, being it in the normal or in the private mode,
the module generates a list of all the entities tracking this page and serving ads on them.

In addition, our system attempts to retrieve the \emph{landing page} of all ads displayed in both modes,
that is, the page of the advertiser that the browser is re-directed to when clicking on its ad~\cite{Kae11LSDM}.
Recall that our system needs the interest category of an ad to make a decision on whether it is profile-based or not.
In order to classify an ad into a topic category, the categorization module (described later in Sec.~\ref{sec:Implementation:Architecture:Components:Categorizer}) requires its landing page.
However, because clicking on every ad to get this information would lead us to commit click fraud~\cite{Jansen07C},
the functionalities provided by our tool in terms of transparency and blocking are limited to those ads where the landing-page information is available without clicking on them.
Despite this limitation, some recent studies~\cite{Kae11LSDM,LiuHotNet13} have reported an availability of the landing page above 80\%.

\paragraph{Categorizer}
\label{sec:Implementation:Architecture:Components:Categorizer}
\noindent
This module classifies the pages visited by the user as well as the landing pages of the ads directed to them, into a predefined set of topic interests. The module employs a 2-level hierarchical taxonomy, composed of 32 top-level categories and 330 bottom-level categories or subcategories. Tables~\ref{tab:TopLevelCategories} and~\ref{tab:SubLevelCategories} show the top-level categories and the subcategories corresponding to three of these categories.

The categorization algorithm integrated into our system is partly inspired by the methodology presented in~\cite{Kae11LSDM} for classifying non-textual ads into interest categories. The algorithm also builds on the taxonomy available at the Firefox Interest Dashboard plug-in~\cite{FID14plugin} developed by Mozilla.

Our categorizer relies on two sources of previously-classified data. First, a list of URLs, or more specifically,
domains and hostnames,
which is consulted to determine the page's category. Secondly, a list of unigrams and bigrams~\cite{Manning99B} that is used when the URL lookup fails. The former type of data is justified by the fact that a relatively small part of the whole Web accounts for the majority of the visits. Also, it is evident that pre-categorized lookup requires few computational resources on the user's browser and can be more precise. The latter kind of information, on the other hand, is justified as a fall-back and allows us to apply common natural-language heuristics to the words available in the URL, title, keywords and content.

For almost each of the top-level categories, the current version of the plug-in incorporates Alexa.com's 500 top Web sites.
Also, the list of URLs includes the pages classified by Mozilla's plug-in (around seven thousand).
On the other hand, the number of English unigrams and bigrams is approximately 76\,000. Three additional lists, although of a fewer number of entries, are also available for French, Spanish and Italian\footnote{Upcoming versions of this Web-browser extension will include more languages.}.
To compile all these words lists, we have built on the following data:

\begin{itemize}
  \item a refined version of the categorization data provided by the Firefox Interest Dashboard extension;
  \item a subset of the English terms available at WordNet 2.0~\cite{Miller95ACMCOM} for which the WordNet Domain Hierarchy~\cite{Magnini00LRE,Bentivogli04MLR} provides a domain label;
  \item a subset of the terms available at the WordNet 3.0 Multilingual Central Repository~\cite{Gonzalez12GW}, to allow the categorization of Web sites written in the aforementioned languages;
  \item and the synset-mapping data between the versions 2.0 and 3.0 of WordNet~\cite{Daude03RANLP}.
\end{itemize}

The categorizer module resorts to these lists only when the hostname and domain are not found in the URL database.
When this happens, the algorithm endeavors to classify the page by using the unigrams and bigrams extracted from the following data fields: URL, title, keywords and content. Depending on the data field in question, the categorizer assigns different weights to the corresponding unigrams and bigrams. In doing so, we can reflect the fact that those terms appearing in the URL, the title, and especially the keywords specified by the publisher (if available), are usually more descriptive and explanatory than those included in the body of the page.

As frequently done in information retrieval and text mining, our Web-page classifier also relies on the term frequency-inverse document frequency (TF-IDF) model~\cite{Salton75COM}. Said otherwise, we weight the resulting category/ies based on the frequency of occurrence of the corresponding unigrams and bigrams, and on a measure of their frequency within the whole Web.

\begin{table*}[htb!]
\centering
\footnotesize
\begin{tabular}{@{}*4l@{}}
\toprule
\multirow{2}{*}{\,\,\quad\quad\quad\quad\quad\quad\quad\quad\quad\textbf{Optimization library}} &  \multicolumn{3}{c}{\textbf{Running time} [s]}\\
  & \,\,\,\,average & \,\,\,variance & maximum \\
  \cmidrule(r){1-1}\cmidrule(l){2-4}
Coin-OR Linear Programming (CLP), v.1.16.6~\cite{CLPsite,Lougee03IBM}& 0.0315505	& 0.0000010	& 0.0460014 \\
GNU Linear Programming Kit (GNULPK), v.4.48~\cite{GLPKsite}	& 0.0337618 &	 0.0000055	 & 0.0681626 \\
Object Oriented Quadratic Programming (OOQP), v.0.99.22~\cite{Gertz03MS}	& 0.0401395 &	 0.0000028	 & 0.0805860 \\
LPSolve, v.5.5.2.0~\cite{Berkelaar04Lpsolve} &	0.0645488	& 0.0000024	& 0.0808482 \\
C Library for Semidefinite Programming (CDSP), v.6.1~\cite{Borchers99OMS}& 0.5878725	& 0.0017888	& 1.1033131 \\
Dual-Scaling Semidefinite Programming (DSDP), v.5.8~\cite{Benson00OP} & 2.0933280	& 0.0137100	& 4.1946620 \\
\rowcolor{Gray}
Coin-OR Interior Point OPTimizer (IPOPT), v.3.12.3~\cite{IPOPTsite,Wachter06MP}& 0.2014676 & 0.1510803	 & 5.7872007\\
\rowcolor{Gray}
Limited Memory Broyden-Fletcher-Goldfarb-Shanno (LBFGS), v.3.0~\cite{Zhu07MS}& 0.2054921	& 0.1669853 & 6.1828331 \\
\rowcolor{Gray}
NLopt, v.2.4.2~\cite{NLOPTsite}& 0.5781220	& 0.0010485	& 0.6520662 \\
 \bottomrule
\end{tabular}
\caption{We tested 6 optimization libraries to compute the solution to the LP problem~\eqref{eq:MinimaxLP}, and another 3 for the divergence-minimization problem~\eqref{eq:uniqueness}. This figure shows the average, the variance and the maximum values of running time, obtained from one thousand problem instances.
Each solver is listed along with the corresponding version number.}
\label{tab:Benchmarking}
\end{table*}

For the sake of computational efficiency, the algorithm stores the categories derived from the user's last 500 visited pages. This way, when the user re-visits one of those pages, the topic categories are obtained directly without needing to go through the process above.

In terms of storage, the whole list of unigrams, bigrams and their corresponding IDF values occupies approximately 1 megabyte in compressed format. We believe this is an acceptable overhead to the plug-in download size.

Lastly, a manual inspection of the categorization results for a large collection of Web pages and ads indicates that the algorithm is, in almost all cases, certainly precise. Further investigation would be required, however, to evaluate the performance of the categorizer in a more rigorous manner.

\paragraph{Optimization Modules}
\label{sec:Implementation:Architecture:Components:Optimization}
\noindent
The optimization modules incorporated in the user side and the server side are responsible for computing the solutions to the problems~\eqref{eq:MinimaxLP} and~\eqref{eq:uniqueness}, and thus obtaining the robust minimax detector and the minimum profile uniqueness, respectively.
The input parameters of the user-side module are the distribution $q$ and the tuples $p^{\textnormal{min}}$ and $p^{\textnormal{max}}$.
On the server side, our system requires the observed clickstream of each ad selector to compute the average profile and the associated uncertainty class.
We would like to remark that the ad transparency and blocking functionalities related to profile uniqueness will only be provided should the user consent to convey such clickstream data.

In the architecture implemented, both modules rely on open-source optimization libraries.
The design of such modules required the examination and comparison of a variety of optimization solvers to this end. Because our system may need to compute the robust detector each time an ad is displayed, we endeavored to prioritize efficiency and reliability on the user side.
These same requirements were also allowed for on the server side.
However, because the minimum-uniqueness values $u_{\textnormal{min}}$ are meant to be computed for \emph{each user}, we opted to lighten the processing and computation in this part of the architecture.
Particularly, instead of processing the profile data every time there is an update on the user side, we specify regular intervals of 1 day (from the time the plug-in is installed) for the exchange of information with the server.
We acknowledge that, depending on the user activity, this might have a certain impact on the accuracy of the profile-uniqueness data provided.

With all these requirements in mind, we performed a benchmark analysis for the LP and divergence-minimization problems. We employed the Matlab optimization toolbox OPTI~\cite{Currie12FCAPO}, and tested one thousand problem instances with random ---although feasible--- values for the inputs mentioned above\footnote{The optimization software was tested on an Intel Xeon E5620 processor, equipped with 8 GB RAM, on a 32-bit Windows 7 operating system.}. For the problem~\eqref{eq:uniqueness}, and when available at the optimization software under test, we also provided the gradient and the Hessian of the objective and constraint functions. In addition, we reduced the complexity of this latter problem by using a top-level representation of $\bar{p}$, $p^{\textnormal{min}}$ and $p^{\textnormal{max}}$ with only 32 categories.

The results are shown in Table~\ref{tab:Benchmarking} for 9 optimization solvers.
Based on our performance analysis, we selected the CLP~\cite{CLPsite,Lougee03IBM} and
IPOPT~\cite{IPOPTsite,Wachter06MP} libraries, which provide a simplex and an interior-point method~\cite{Boyd04B}, respectively.
The two solvers exhibited the lowest average running time in our analysis, with 32 and 201 milliseconds respectively, as well as acceptable variance values.
It is worth mentioning that all problem instances were solved satisfactorily by the libraries tested,
and that the two solvers chosen are available under the Eclipse Public License~\cite{EclipseLicenseV1}.

In our system, both solvers were configured to have a maximum allowable running time.
When our extension is installed for the first time, it runs several problem instances to set this parameter;
this is for the computation of the robust minimax interest-based ad detector.
On the server side, the computation of the minimum value of user-profile uniqueness is limited to 0.5 seconds.

\paragraph{Blocking Policies}
\label{sec:Implementation:Architecture:Components:Policy}
\noindent
The functionality of this module is to apply the ad-blocking policies defined by the user.
Its current implementation simplifies the formal policy notation presented in Sec.~\ref{sec:Implementation:Functionalities:Examples}, in an attempt to provide an easy-to-use interface and thus enhance usability.

\begin{figure}
\centering
\includegraphics[scale=0.60]{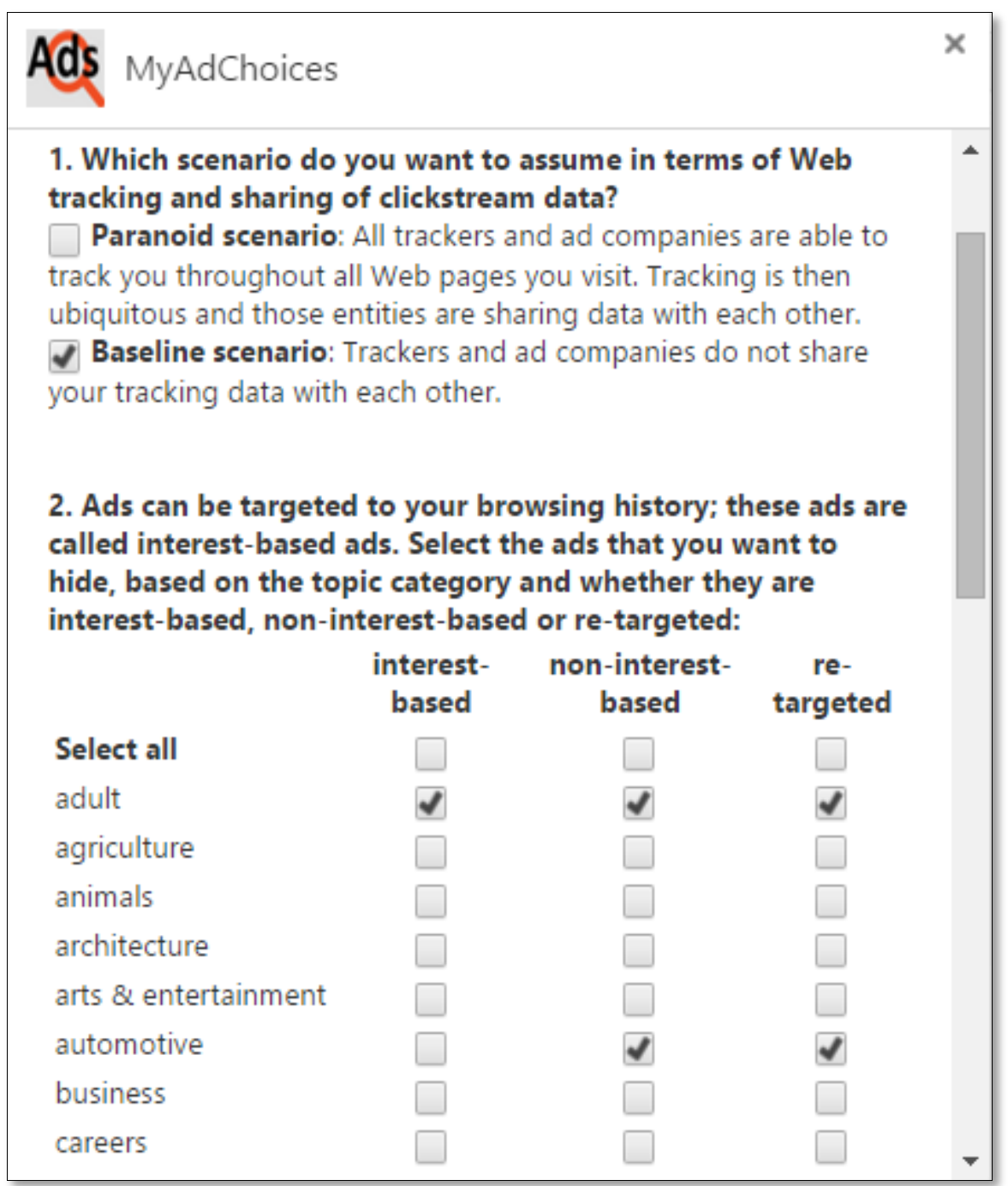}
\caption{The configuration panel shown in this figure allows users to define fine-grained, ad-blocking policies. The options available to users include filtering out ads per interest category, behavioral and retargeting advertising. Although not displayed in this figure, users can also denote ad-blocking conditions depending on the uniqueness of the profiles that ad selectors might potentially build.}
\label{fig:PolicyPanel}
\end{figure}

With this aim, our extension allows users to define policies \emph{only} with negative $\ensuremath{\mathit{sign}}$.
That is, instead of specifying which ads should be displayed (+) and which ones should be blocked (-),
we just enable the latter blocking declaration, which may facilitate the definition of such policies.
In addition, the specification of percentile values of profile-uniqueness is, in this implementation, reduced to a binary choice:
users can only decide if they wish to block (or allow) those entities which may have compiled ``very unique'' profiles of them,
meaning that $\pi_{u_{\textnormal{min}}} \geqslant 90\%$.
Fig.~\ref{fig:PolicyPanel} shows the configuration panel by which users may configure blocking policies, as well as the scenario they wish to assume in terms of Web tracking.

The operation of this module is described next.
When a user visits a page, the module waits for the categorizer to send the topic category of each ad to be displayed.
Then, it receives the robust minimax interest-based ad detectors of each of the entities delivering those ads.
And finally, it consults an internal database (i.e., on the user side) to obtain the minimum uniqueness values associated with such entities.
With all this information, our system only needs to verify if each ad constraint is satisfied and, accordingly, decide whether to block the ad or not.

\begin{figure}
\centering
\includegraphics[width=\columnwidth]{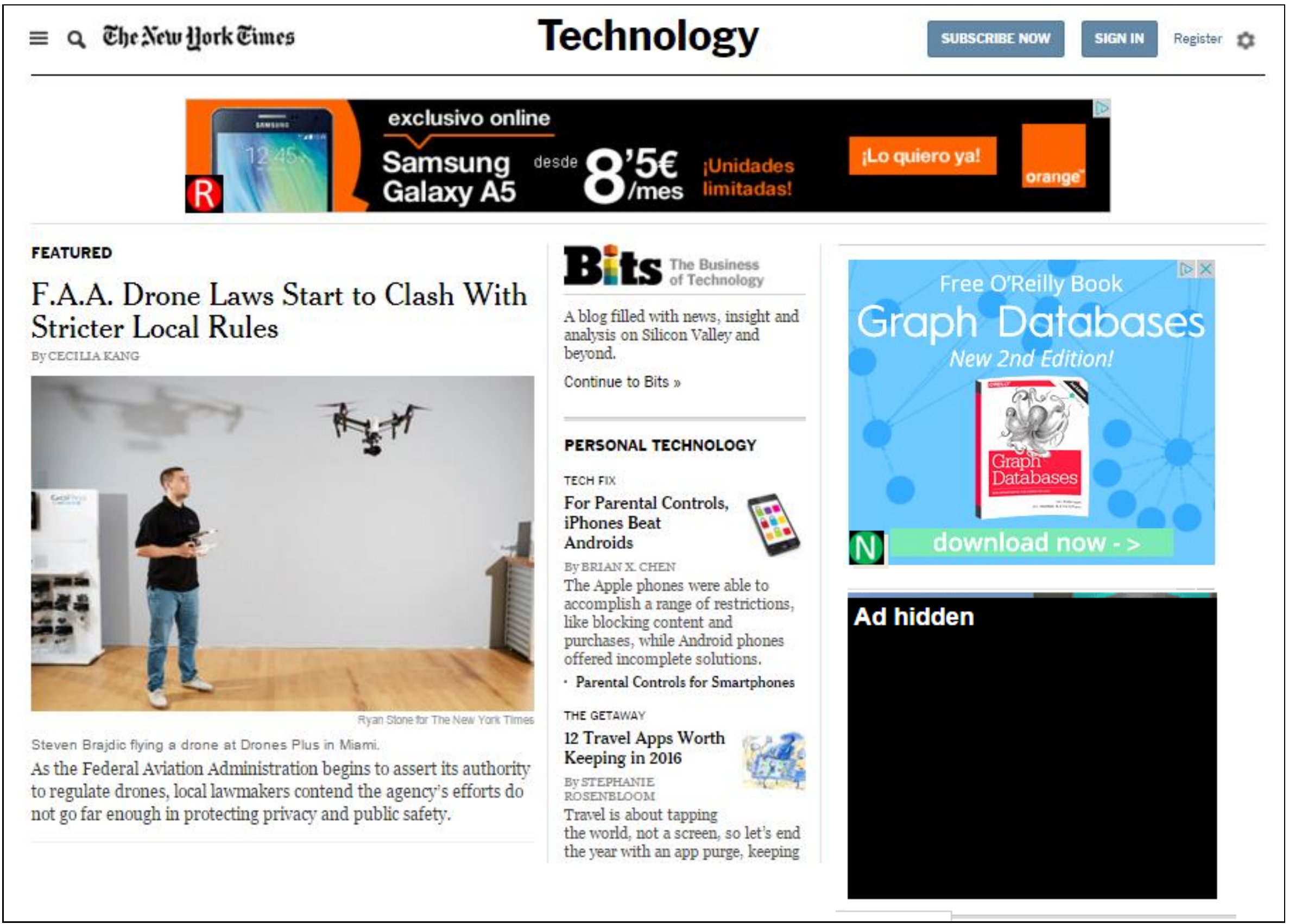}
\caption{We show a screenshot of the ads identified by our system in The New York Times' Web site.
One of these ads is classified as retargeted, another as non-interest-based, and the bottom-right one is hidden according to the user's blocking policy.}
\label{fig:Masks1}
\end{figure}

We must highlight that our system does not block the ads in the same sense as current ad-blocking technologies do.
While these technologies prevent third-party network requests\footnote{AdBlock Plus~\cite{AdblockPlus15Web}, for example, do not block \emph{all} third-party network requests but only those blacklisted~\cite{AdblockPlus15Lists}.} from being sent, our Web-browser extension does allow them.
It is only when the page is completely loaded and thus the ads (if any) are displayed, that our system decides to hide them or not by applying a black mask on top of them\footnote{On a technical note, the system might alternatively remove the ad image.}.
To highlight this particular aspect, we refer to the action of blocking more precisely as \emph{hiding} or \emph{obfuscation}.
Fig.~\ref{fig:Masks1} shows a screenshot of the ads processed by our tool in a particular Web page.

The tool notifies users about the kind of ads received through a small icon placed on the left corner of each detected ad.
The icons indicate if an ad is interest-based (red), retargeted (red), non-interested-based (green), it is blocked according to the user's policy (black), or the system cannot make a decision (orange). This latter case occurs, for example, when the ad's landing page is not available or the categorizer cannot classify it;
when there is insufficient data to train the PMF models of $p$ and $q$; or when the execution of the optimization solver exceeds the maximum allowable running time.

\section{Evaluation}
\label{sec:Evaluation}
\noindent
In this section, we empirically evaluate the proposed system and analyze several aspects of behavioral advertising.
The analysis of this form of advertising is conducted from the ads as well as browsing data of 40 users of \emph{MyAdChoices}.
To the best of our knowledge, this study constitutes the first, albeit preliminary, attempt to investigate behavioral targeting and profile uniqueness in a real environment from real user browsing profiles.

\subsection{Data Set}
\label{sec:Evaluation:DataSet}
\noindent
We distributed \emph{MyAdChoices} to colleagues and friends and asked them to install it and browse the Web normally for one month.
The experiment was conducted from December 2015 to January 2016.
The data collected by our Web-browser extension were sent to our servers every one hour.
On the other hand, the extension was configured for a fraction of revisited pages of 100\%.
That is, every page browsed by a user was revisited by our system in the incognito mode.

The participants were mostly researchers and students based in our countries of residence, France, India and Spain.
No attempt was made to link the gathered data to the personal identities of the volunteers.
As a preprocessing step, we removed those users who visited less than 100 sites,
leaving a total of 40 users.

\subsection{Results}
\label{sec:Evaluation:Results}
\noindent

\subsubsection{System Performance}
\label{sec:Evaluation:Results:Performance}
\noindent
Evaluating an ad-transparency tool is extremely challenging since the ground truth of targeting decisions is unknown.
The effectiveness of these tools has been occasionally assessed through manual inspection~\cite{Lecuyer14SSYM,Datta15PET}.
However, this approach has been recently shown to be extremely prone to errors~\cite{Lecuyer15CCS}.
In this section, we evaluate the error probability of the interest-based ad detector
bearing in mind
the impossibility of checking a detector's decisions with the true condition of the tested ads (i.e., whether they are actually interest-based or not).

\begin{figure}
\centering
\includegraphics[width=\columnwidth]{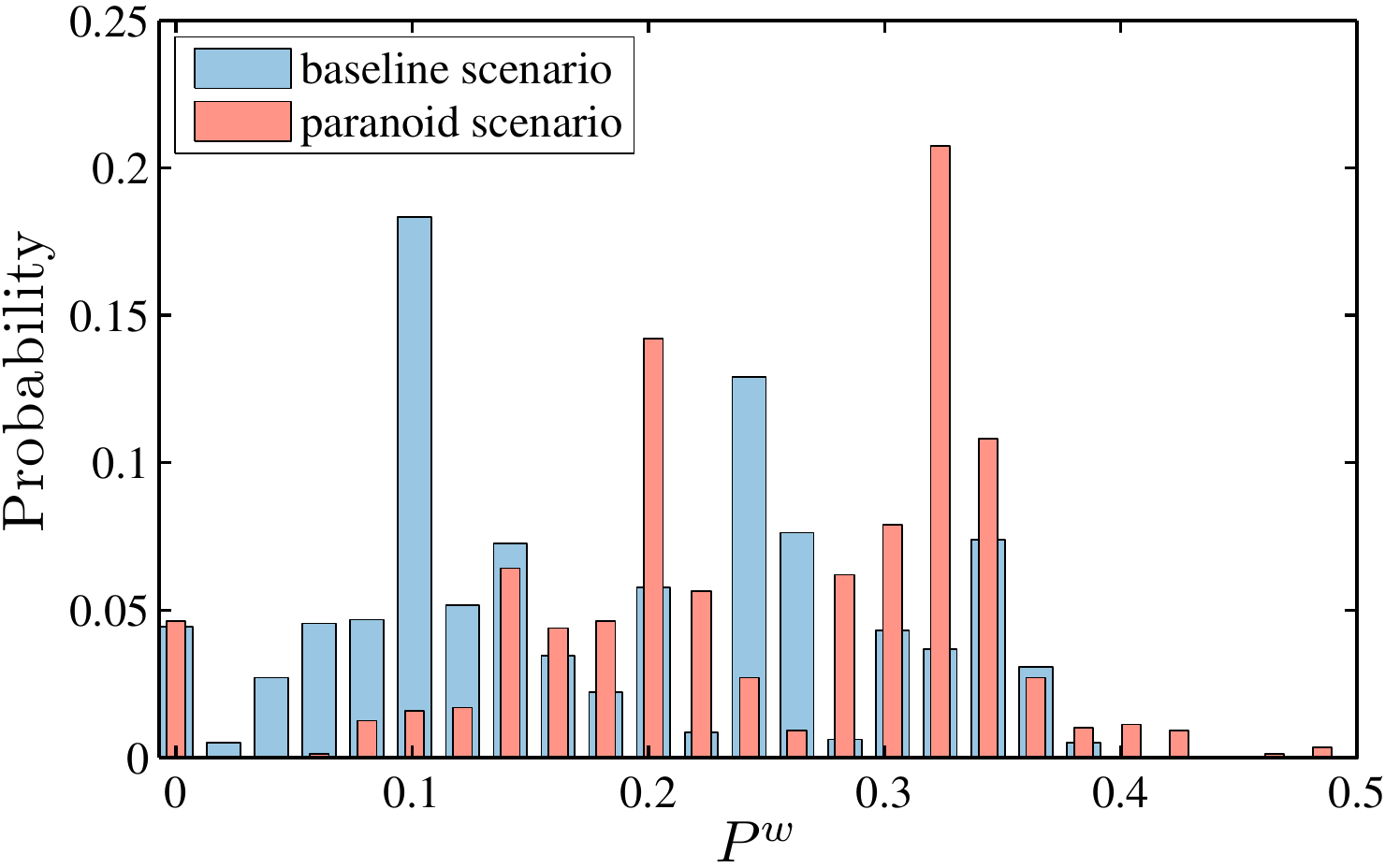}
\caption{PMF of the worst-case error probability for the two scenarios assumed in this work.}
\label{fig:Fig1}
\end{figure}

Before proceeding with this evaluation, we first report the availability of categorization data in our data set.
Recall that our system classifies ads into topic categories from their landing pages. To this end, the categorization module makes use of the words included in the landing page's URL, keywords, title and content.
In our series of experiments, we found that just 0.60\% of ads could not be categorized by using this information,
which represents a good availability index.
In most of the cases, the reason was the lack of language support.
As explained in Sec.~\ref{sec:Implementation:Architecture:Components:Categorizer}, currently our categorization module works only for English, French, Spanish and Italian.

\begin{figure*}[htb!]
\centering
\includegraphics[scale=0.55]{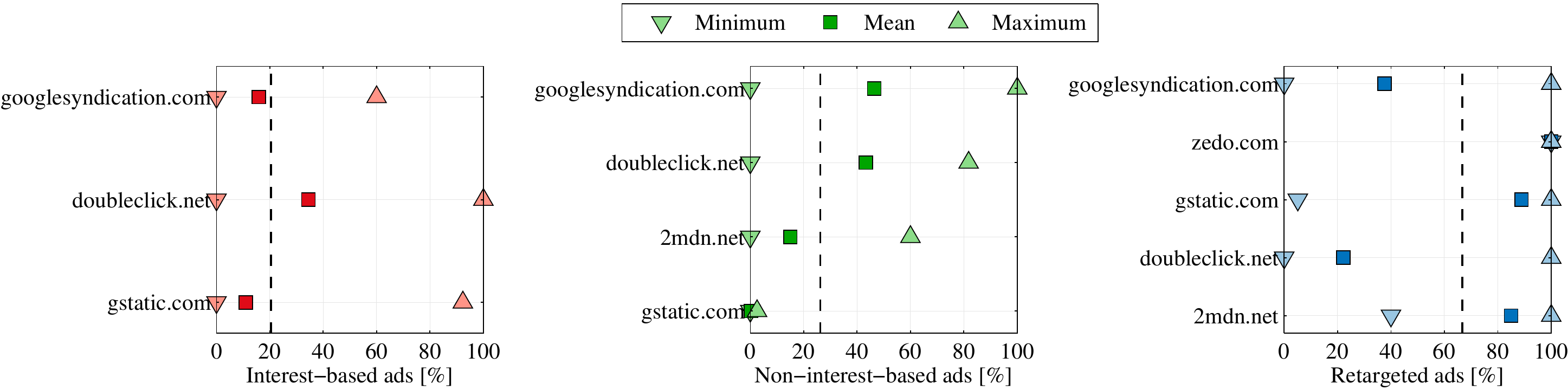}
\caption{\emph{Ad selectors}. Interest-based, non-interest-based and retargeted ads for the \emph{baseline} scenario.}
\label{fig:Fig2Baseline}
\end{figure*}
\begin{figure*}[htb!]
\centering
\includegraphics[scale=0.45]{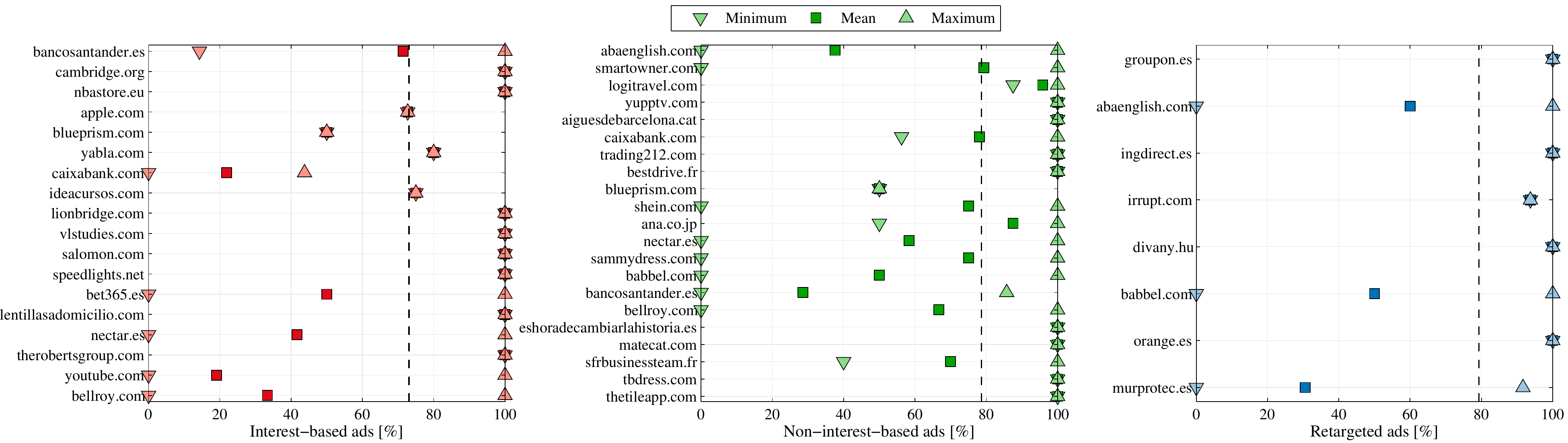}
\caption{\emph{Advertisers}. Interest-based, non-interest-based and retargeted ads for the \emph{baseline} scenario.}
\label{fig:Fig3Baseline}
\end{figure*}

Having checked the performance of our categorizer, now we turn to the robust minimax detector.
In all the executions of the optimization library CLP (including both the baseline and the paranoid scenarios) no single error was reported to our servers. That is, our system was able to successfully compute said detector, without exceeding the maximum allowable running time for this computation, set to 0.5 seconds in these experiments. Likewise, the IPOPT software did not report any error when computing the values of minimum uniqueness.

Fig.~\ref{fig:Fig1} shows the PMF of the probability of error of the interest-based ad detector.
In the baseline scenario, we observe a mean and a variance of 0.1827 and 0.0105, respectively. In the paranoid case, these two moments yield 0.2504 and 0.0094.
Two remarks are in order from these figures.
First, both cases exhibit relatively low error probabilities, with expected values roughly lower than 1/4.
Secondly, the paranoid scenario seems to be slightly more prone to errors in terms of interest-based ad detection.
One possible explanation for this is a greater semblance between the distributions $p$ and $q$ in this scenario.
Intuitively, the more dissimilar these distributions are, the lower is the probability of incorrectly identifying an interest-based ad.

\begin{table}[htb!]
\label{tab:BTResults}
\centering
\scriptsize
\caption{Minimum, mean and maximum percentage values of interest-based, non-interest-based and retargeted ads over all users in our data set.}
\label{tab:BTResults}
\begin{tabular}{lcccccc}
      & \multicolumn{3}{c}{\textbf{Baseline scenario} [\%]}                                                   & \multicolumn{3}{c}{\textbf{Paranoid scenario} [\%]}                                                    \\      \cmidrule(lr){2-4} \cmidrule(lr){5-7}
      & \multicolumn{1}{c}{min.} & \multicolumn{1}{c}{mean} & \multicolumn{1}{c}{max.} & \multicolumn{1}{c}{min.} & \multicolumn{1}{c}{mean} & \multicolumn{1}{c}{max.} \\ \cmidrule(lr){1-4} \cmidrule(lr){5-7}
\textbf{Interest-based}               & 0      & 13.2     & 60.0      & 0      &17.8      & 66.7     \\ \cmidrule(lr){1-5} \cmidrule(lr){5-7}
\textbf{Non-interest-based}       & 0      &31.7      & 78.4      & 0      & 29.4     & 76.1 \\ \cmidrule(lr){1-5} \cmidrule(lr){5-7}
\textbf{Retargeted}                     & 0      & 55.1    & 100      & 0      & 52.8      & 100      \\ \cmidrule(lr){1-5} \cmidrule(lr){5-7}
\end{tabular}
\end{table}

\subsubsection{Behavioral and Retargeted Advertising}
\label{sec:Evaluation:Results:Behavioral}
\noindent
This section examines several aspects of behavioral advertising and retargeting,
including an analysis of the entities delivering such forms of advertising; the topic categories most targeted in our experiments;
the discrepancy between the baseline and paranoid scenarios; and a preliminary study of the relationship between interest-based advertising and profile uniqueness.

Some general figures on behavioral and retargeted advertising are shown in Table.~\ref{tab:BTResults}.
To obtain these figures, we computed, for each user with a minimum of 10 ads received, the percentage of interest-based, non-interest-based and retargeted ads.
The minimum, mean and maximum values of those percentages over all users are the values represented in this table.

The results clearly indicate that retargeting is the most common ad-targeting strategy, followed by non-interest-based advertising and behavioral targeting.
This order is observed both in the baseline and in the paranoid scenario, with small differences
in the percentage values.
One of the most interesting results is the relatively small prevalence of behavioral targeting, which accounts for one third of retargeted ads.
This is in contrast with previous work reporting higher average percentages of this type of advertising for fake profiles~\cite{Carrascosa15CoNEXT},
but in line with recent marketing studies~\cite{Allen14TR} which point out that retargeted ads are preferred to interest-based ads
in a proportion 3:1.

\begin{figure*}[htb!]
\centering
\includegraphics[scale=0.55, angle=90]{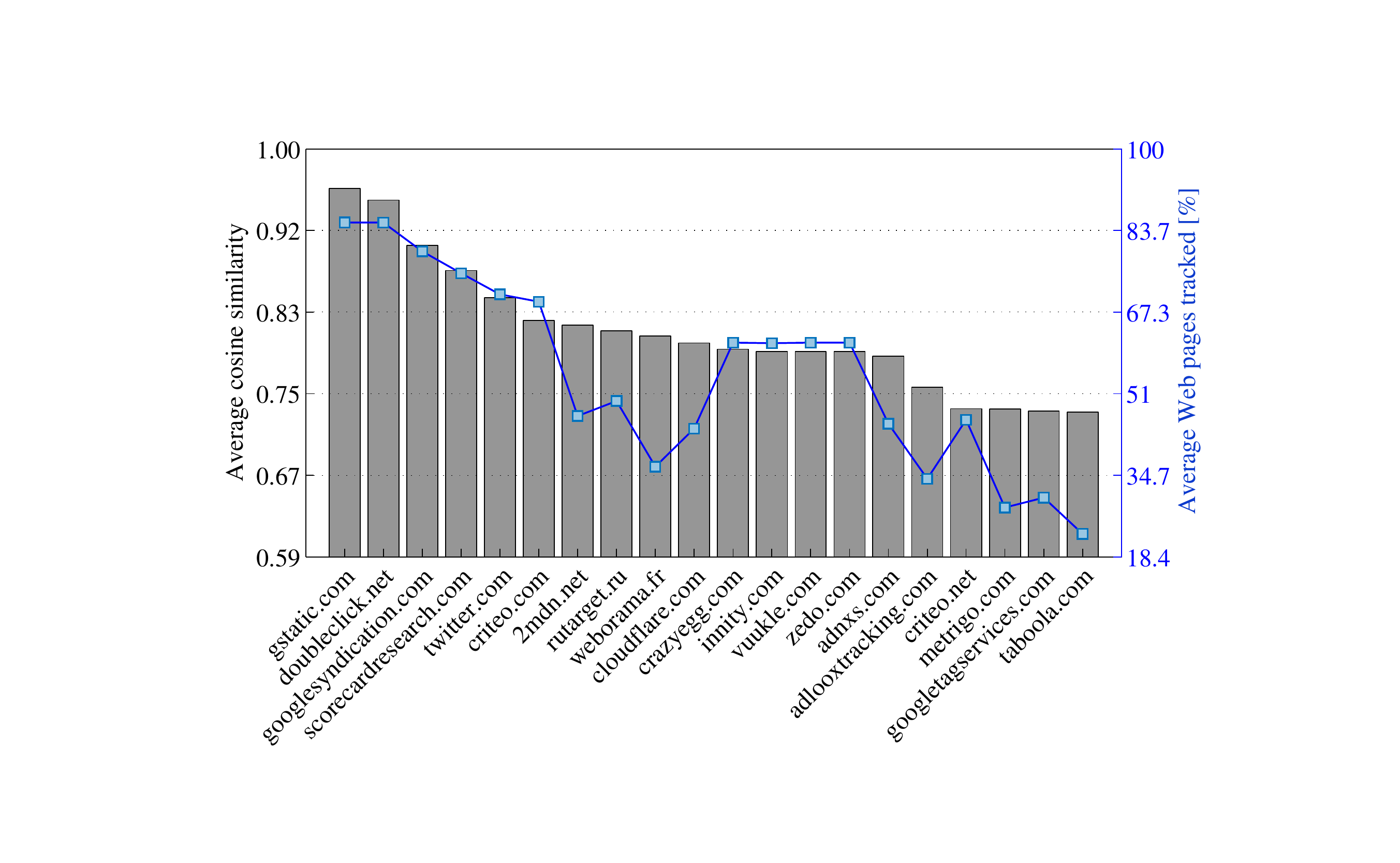}
\caption{We show the cosine-similarity values between the actual and the observed profiles, averaged over all users and per tracking entity.}
\label{fig:CosineSimilarity}
\end{figure*}

\begin{figure*}[htb!]
\centering
\includegraphics[scale=0.55]{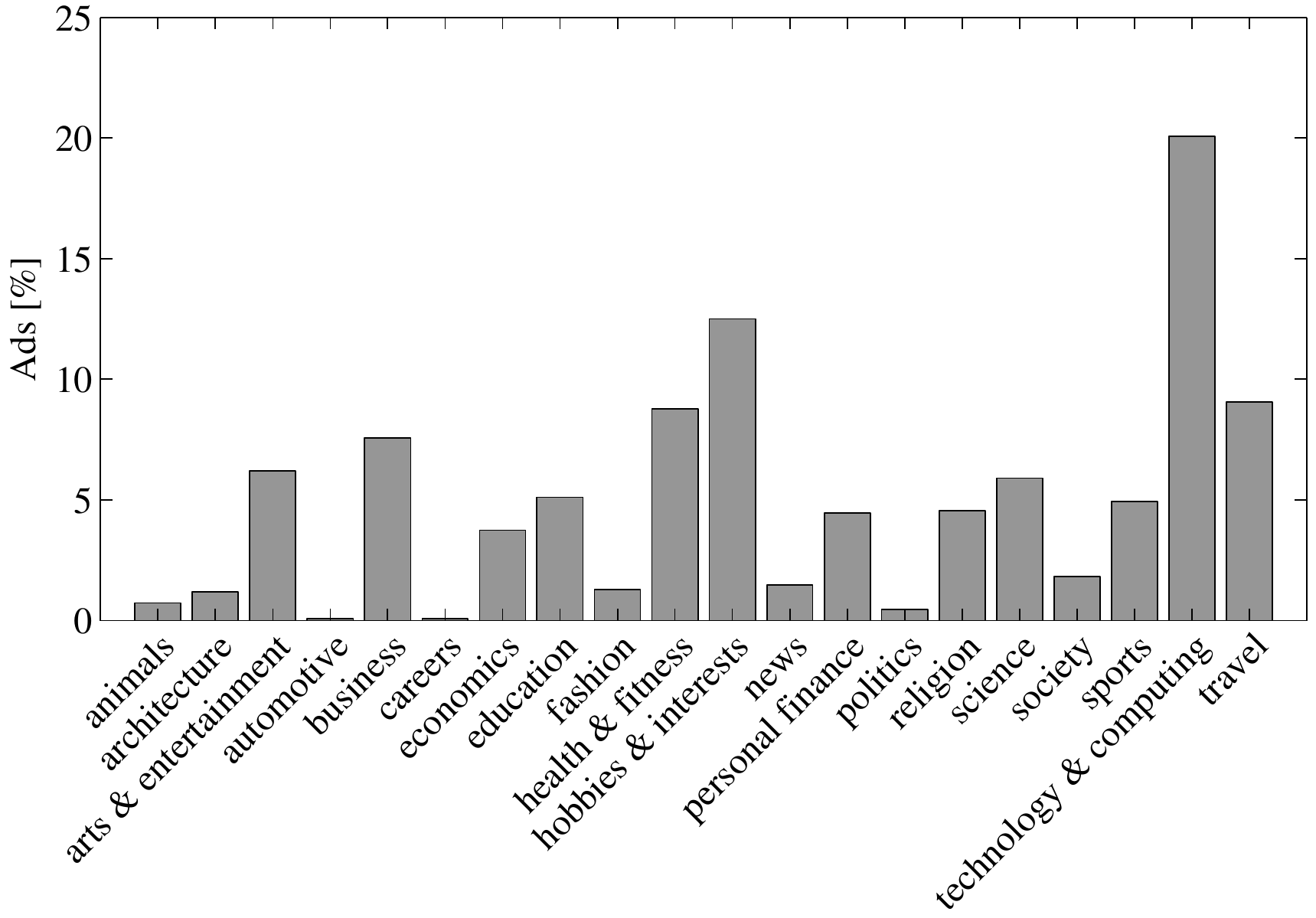}
\caption{Percentage of ads across the top 20 topic categories.}
\label{fig:TopCatsPMF}
\end{figure*}

\paragraph{Ad Selectors and Advertisers}
\label{sec:Evaluation:Results:Behavioral:Entities}
\noindent
In this subsection, we examine the ad selectors which, in our data set, were responsible for the delivery of behavioral, non-behavioral and retargeted advertising.
We computed, to this end, the percentage of interest-based, non-interest-based and retargeted ads served by each of these entities.
Fig.~\ref{fig:Fig2Baseline} depicts the minimum, mean and maximum values of such percentages for each ad selector delivering a minimum of 10 ads; these results correspond to the baseline scenario.
In each of the three diagrams, ad selectors were sorted in decreasing order of total number of served ads, from top to bottom.
The dot vertical lines indicate average percentages over the ad selectors displayed.

The figure in question shows only five ad selectors. In our data set, these entities were responsible for 98.99\% of the total number of ads.
Not entirely unexpectedly, Google's ad companies (\texttt{googlesyndication.com}, \texttt{doubleclick.net} and \texttt{gstatic.com}) were the ones monopolizing the three ad classes.
The former ad platform was observed to target mostly non-interest-based and retargeted ads,
whereas DoubleClick and \texttt{gstatic.com} focused on behavioral advertising and retargeting, respectively.
The remaining ad selectors were \texttt{zedo.com} and \texttt{2mdn.net}. The majority of ads served by these ad companies were retargeted.
Lastly, the paranoid case exhibits similar results and is omitted for the sake of brevity.

The same methodology was used to analyze the advertisers of our data set, and to generate Fig.~\ref{fig:Fig3Baseline}.
This figure shows Banco Santander, Cambridge University Press, NBA Store and Apple as the advertisers with the highest rates of behavioral advertising.
SmartOwner, Logitravel.com, YuppTV and CaixaBank, on the other hand, lead the ranking of non-interest-based ads,
and Groupon, ABA English and Ing Direct are the companies most interested in retargeting.
Although we cannot derive a general rule from these results, we note that large companies are more frequent in the behavioral-targeting list than in that of non-interest-based ads. This might be an immediate consequence of the higher chances of such firms ---for example--- to win ad-auctions at RTB, compared to companies with limited purchasing power.

\begin{figure*}[htb!]
\centering
\includegraphics[scale=0.60]{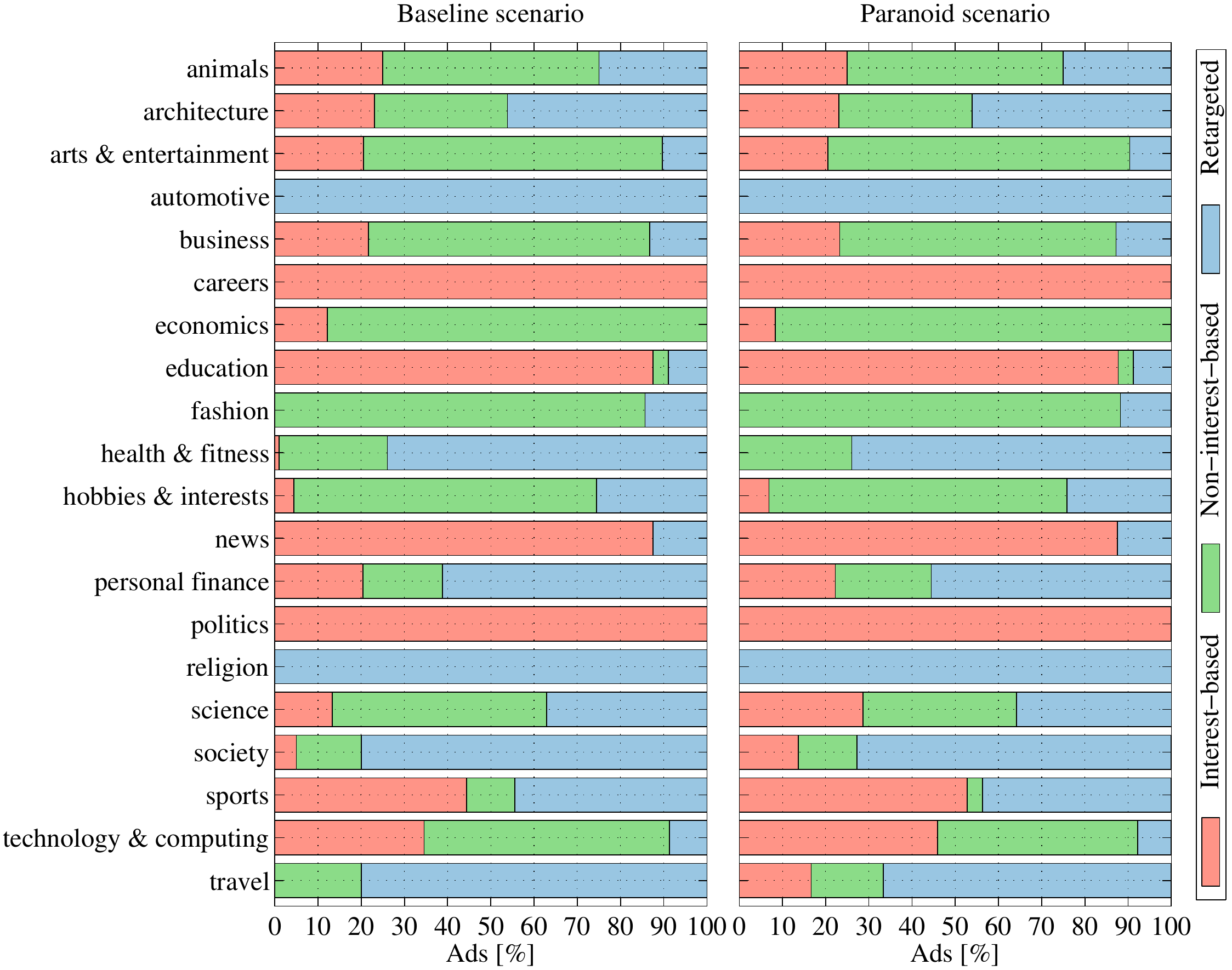}
\caption{Some of the top-level interest-categories targeted in the baseline and the paranoid scenarios.}
\label{fig:TopCats}
\end{figure*}

\paragraph{Baseline and Paranoid Scenarios}
\label{sec:Evaluation:Results:Behavioral:BaselineParanoid}
\noindent
Next, we analyze the overall percentage of coincidence between the baseline and the paranoid scenarios in terms of interest-based ad detection.
To this end, for each ad we checked if the decision made by the detector in the baseline mode matched the decision made by the detector in the paranoid case.

The percentage of matching observed in our data set was certainly high, especially for the ad platform \texttt{gstatic.com}, which yielded 97.4\%.
Although smaller, the percentages of coincidence for DoubleClick (75.6\%) and \texttt{googlesyndication.com} (87.0\%) were also remarkable.
A plausible explanation to this behavior is the semblance of the profile $p$ estimated in both scenarios,
which might indicate that \texttt{gstatic.com} relied only on its own tracking data and thus did not enrich this information with browsing profiles from other sources.

Precisely, the semblance of the profiles $p$ and $t$ is investigated in our next figure, Fig.~\ref{fig:CosineSimilarity}.
Recall that these profiles are estimated from the observed and the actual clickstreams respectively.
To compute Fig.~\ref{fig:CosineSimilarity}, we kept a record of all entities tracking users' visits;
these entities were ad platforms, advertisers and also data-analytic trackers.
Then, from said records, we calculated the percentage of pages tracked by each of these entities, as well as the cosine similarity between the observed and actual profiles. The figure at hand shows these percentage and similarity values averaged over all users.

A couple of remarks follow from this figure.
First, Google's ad platforms are the entities with the most extensive tracking capabilities.
Particularly, \texttt{gstatic.com}, DoubleClick and \texttt{googlesyndication.com} tracked users on 92.9\%, 88.2\% and 81.2\% of the visited pages.
An immediate consequence of this are the high values of cosine similarity observed.
Secondly, the results are consistent with the percentages of scenario matching provided at the beginning of this subsection.
Thirdly, the profiles $p$ of ad companies with limited tracking capabilities like Metrigo and Taboola were observed to be relatively similar to the corresponding actual profiles. Although it is not possible to find an accurate answer for this result, the reason might be found in the model of user profile based on \emph{relative} frequencies.

Finally, we would like to emphasize the appropriateness of the proposed scenarios for the particular ad selectors examined in these experiments. Recall that the baseline scenario does not contemplate the sharing of tracking information with other ad selectors and trackers, whereas the paranoid case does; this latter scenario also considers that tracking is ubiquitous.
The results provided throughout this experimental section build on the assumption that \texttt{googlesyndication.com}, DoubleClick and \texttt{gstatic.com} operate independently in the baseline scenario.
However, since they are all Google ad companies, one might expect that these three firms would have exchanged information with each other.
The paranoid scenario precisely captures this possible exchange of tracking data.
Also, the ubiquitousness of tracking is justified by the fact that these ad platforms combine for a total of 99.08\% pages tracked (i.e., they track users almost in all pages they visit).

\paragraph{Interest-Categories Targeted}
\label{sec:Evaluation:Results:Behavioral:Cats}
Fig.~\ref{fig:TopCatsPMF} plots the probability distribution of the ad-topic categories.
In this figure, we considered only those topics for which we collected a minimum of 5 ads.
The results indicate that the most popular interest categories were ``technology \& computing'', ``hobbies \& interests'', ``travel'' and ``health \& fitness'',
with percentages of 18.4\%, 11.5\%, 8.3\% and 8.1\%, respectively.

Fig.~\ref{fig:TopCats} illustrates, on the other hand, the targeting strategies that were observed in each of the 20 categories represented in Fig.~\ref{fig:TopCatsPMF}.
As can be seen, very similar results were reported for the baseline and the paranoid scenarios.
Our findings show that retargeted ads were more frequent on categories like ``automotive'', ``religion'', ``society'' and ``travel'',
which seems to be partly in accordance with some marketing surveys~\cite{Freed12TR,Butler10TR}.
On the other hand,
profile-based ads were observed more predominantly on ``careers'', ``education'', ``news'' and ``politics'',
and non-interest-based ads were largely targeted to ``fashion'', ``economics'' and ``hobbies \& interests''.

\begin{figure}
\centering
\includegraphics[width=\columnwidth]{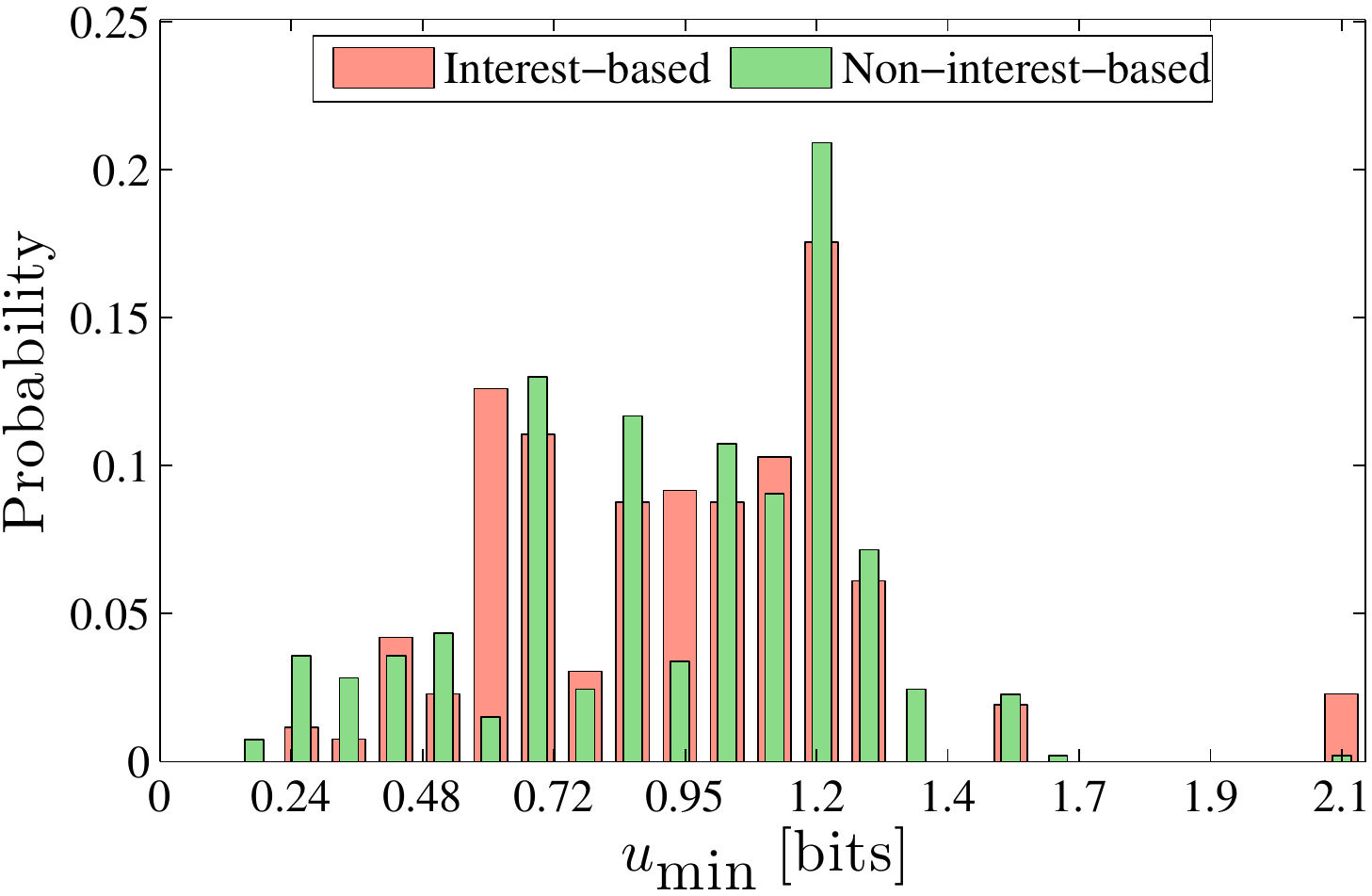}
\caption{We show the PMFs of the profile-uniqueness values analyzed, when ads are classified as interest-based and when they are considered to be non-interested-based.}
\label{fig:Uniqueness}
\end{figure}

\paragraph{Behavioral Targeting and Profile Uniqueness}
\label{sec:Evaluation:Results:Behavioral:Unique}
\noindent
In our last experiments, we briefly explore whether common browsing profiles are more likely (or not) to receive interest-based ads.
With this purpose, for each ad classified as interest-based and non-interest-based, we analyzed the minimum-uniqueness values of the ad selector serving it.
The probability distributions of such values are plotted in Fig.~\ref{fig:Uniqueness}.

As can be observed, the two PMFs are very similar, which clearly means that the probability of delivering an interest-based ad may not depend on the uniqueness of the observed profile.
In fact, the expected values of these distributions are 0.8949 bits for profile-based ads, and 0.8834 bits for non-interest-based ads;
and the KL divergence (a measure of their discrepancy) yields 0.4344 bits.
On the basis of the evidence currently available, it seems fair to suggest that the uniqueness or commonality of a profile is not a feature that ad selectors in general use to decide their user-targeting strategies.
Further evidence supporting this assertion, however, would require the analysis of larger volumes of data.

\section{Related Work}
\label{sec:RelatedWork}
\noindent
This section reviews the state of the art relevant to this work.
We proceed by exploring, first, the current software technologies aimed at blocking ads;
and secondly, we examine those approaches intended to provide transparency to online advertising.

\subsection{Ad Blockers}
\label{sec:RelatedWork:Blockers}
\noindent
The Internet abounds with examples of ad-blocking technologies.
In essence, these technologies act as firewalls between the user's Web browser on the one hand, and the ad platforms and tracking companies on the other.
Specifically, ad blockers operate by preventing those HTTP requests which are made when the browser loads a Web page, and which are not originated by its publisher.
These requests are commonly referred to as third-party network requests, as mentioned in the introductory section of this work.

Most of these tools are implemented as open-source browser plug-ins, and carry out said blocking with the help of a data base or \emph{blacklist} of ad platforms and trackers.
Basically, these lists include regular expressions and rules to filter out the third-party network requests that are considered to belong to ads or trackers. The maintenance of such blacklists is done manually by the technologies' developers and in some cases by user communities.
Some of the most popular ad-blockers are Adblock Plus~\cite{AdblockPlus15Web} and Adblock~\cite{Adblock15Web}.
Within this list of blocking technologies, we also include anti-tracking tools like Ghostery~\cite{GhosteryWeb}, Disconnect~\cite{Disconnect15Web}, Lightbeam~\cite{Lightbeam15Web} and Privacy Badger~\cite{PrivacyBadgerEFF},
which, from an operational point of view, work exactly as ad blockers and thus may as well block ads.

A middle-ground approach for ad-blocking has recently emerged that uses whitelists to allow only ``acceptable ads''.
The criteria for acceptability typically comprise non-invasiveness, silence and small size~\cite{Sayer15PCWorld}.
However, because these criteria ultimately depend on the ad blockers' developers,
this approach does not signify any real advance in the direction of returning users control over advertising.
Indeed, this ``acceptable-ads'' approach has caused a great controversy in the industry,
when it came to the public that the most popular ad blocker
was accepting money from some of the whitelisted companies~\cite{Cookson15FT}.

\begin{table*}[htb!]
\caption[MyAdChoices versus other behavioral-targeting transparency tools]%
{Comparison between MyAdChoices and other tools that may provide transparency to behavioral advertising.}
\label{tab:Comparison}
\centering
\footnotesize
\begin{tabular}{p{2.5cm}p{2.5cm}p{10cm}}
\toprule
\textbf{Approaches} & \textbf{Type of tool} & \textbf{Disadvantages} \\ [0.8ex]
\toprule
\cite{Carrascosa15CoNEXT,Carrascosa14arXiv}
& research platform
& $\circ$ valid for single-category profiles,\\
& & $\circ$ transparency functionality available only on weather pages,\\
& & $\circ$ inaccurate model of the ad-delivery process, \\
& & $\circ$ parallel browsing in incognito mode, \\
& & $\circ$ only paranoid scenario, \\
& & $\circ$ multiple user-targeting objectives not allowed; \\\midrule

\cite{Barford14WWW}
& research platform
& $\circ$ valid for single-category profiles,\\
& & $\circ$ transparency functionality limited to users visiting the same pages,\\
& & $\circ$ generic and contextual ads are omitted, \\
& & $\circ$ only paranoid scenario; \\\midrule

\cite{LiuHotNet13}
& research platform
& $\circ$ only for DoubleClick,\\
& & $\circ$ simplified model of the ad-delivery process (e.g., generic ads and RTB ignored, only for long-term user profiles),\\
& & $\circ$ binary decision, i.e., ads are either contextual or interest-based, \\
& & $\circ$ inconsistent model of tracking and sharing of user data; \\\midrule

\cite{Lecuyer14SSYM,Lecuyer15CCS,Datta15PET}
& research platform
& $\circ$ not scalable for Web browsing~\cite{Lecuyer14SSYM,Lecuyer15CCS},\\
& & $\circ$ unacceptable network traffic and computational overhead, if intended for end users~\cite{Lecuyer14SSYM,Lecuyer15CCS},\\
& & $\circ$ may detect retargeting but not behavioral advertising; \\\midrule

MyAdChoices & end-users tool & $\circ$ a fraction of revisits in incognito mode. \\ \bottomrule
\end{tabular}
\end{table*}

\subsection{Advertising Transparency}
\label{sec:RelatedWork:Transparency}
\noindent
To the best of our knowledge, in terms of transparency, our work is the first to provide \emph{end-users} with detailed information about behavioral advertising in \emph{real-time}.
As we shall see next,
only a couple of previous works tackle the problem of interest-based ad detection.
The major disadvantage of these few existing approaches,
however, is that they are not intended for end-users,
i.e., they are not designed to be used by a single user who wishes to find out what particular ads are targeted to them.
Instead, these approaches consists in platforms
aimed at collecting and analyzing advertising data at large scale for \emph{research} purposes.
In general,
they allow running experiments in a limited and controlled environment,
and studying the ads displayed to very specific and artificially-generated user profiles.

In this subsection we shall examine these works, bearing in mind that
none of them are conceived as a tool that users can directly and fully benefit from it.
In addition, and equally importantly, we shall see that these proposals rely on a too simplistic, and in many cases erroneous, model of the actual ad-delivery process.
Also, they very often resort to simple heuristics, not rigorously justified, to conduct their measurement studies on behavioral advertising.

In contrast to these works, we propose a formal study of this form of advertising that builds on a more general, accurate model of the ad-serving process, which takes into account its complexity and the new paradigm of RTB, and which addresses the challenges others simply neglected.
We proceed by following a mathematically grounded methodology that capitalizes on the fields of statistical estimation and robust optimization.
Besides, compared to these works, our analysis of behavioral targeting does not only determine if an ad is interest-based or not,
but also
it explores a crucial aspect of the interests tracked and profiled by ad companies, namely, the commonality of user profiles.
Next, we elaborate more on these proposals.

The first attempt to identify the challenges that may arise when measuring different aspects of online advertising was done in~\cite{Guha10IMC}.
Although not particularly interested in behavioral targeting,
the authors investigated aspects like
the impact of page reloading and cookies on advertising,
and highlighted the difficulties found through some simple experiments.

Following this work, \cite{Carrascosa15CoNEXT,Carrascosa14arXiv}~proposed a platform which automatizes the collection of certain statistics about behavioral targeting. The proposed platform creates artificial user profiles with very specific, non-overlapping topic categories (i.e., profiles with active categories only in sports, only travel, and so on) by emulating the visits to pages related to those topics. The tool in question alternates this training browsing with visits to weather Web pages,
where they check if the categories of the received ads match the category of the corresponding profile; the authors justify the use of these weather-related pages by arguing that, there, contextual ads are detected more easily.
To carry out this checking, the tool first filters out those landing pages which may correspond to generic and contextual ads.
With this aim ---and similarly to our tool---, it revisits, in incognito mode, each visit to a weather page and keeps a record of the ads delivered in this session.
By eliminating the landing pages common to both sessions, the authors claim to discard the \emph{majority} of untargeted and content-based ads.

Apart for the fact that said platform is not intended for end-users nor provides real-time ad-transparency functionalities, the most important drawback is its extremely limited scope of application. First, it only works for single-interest profiles, and secondly, transparency can only be brought in such weather pages, which provides very simplistic and superficial insight into behavioral targeting.
Nonetheless, this is not the only limitation. To detect interest-based ads, the authors make the mistake of oversimplifying the ad-delivery process by assuming some sort of determinism: they consider that \emph{most} of the non-interest-based ads a user may receive in a tracked and in a free-tracked session will be exactly the same, which totally neglects the inherent randomness of the ad-serving process.
Besides, the authors do not consider the particular ad platform serving an ad and therefore implicitly assume ---although they do not mention it--- a worst-case or paranoid scenario in terms of tracking and sharing of data. This is in contrast to our work, which in addition considers a baseline scenario for tracking.

Finally, the cited works~\cite{Carrascosa15CoNEXT,Carrascosa14arXiv} evaluate their approach by using a distance measure between the terms appearing in the ads' landing pages and those in the training pages. While this quantifies the similarity between the ads' topic categories and profiles' single categories,
the authors do not assess the method to detect profile-based ads.
An important consequence of this lack of evaluation is that generic ads belonging to the profile's active category will always be classified as interest-based (provided that they have not been delivered in the incognito sessions), and the platform will not report any error on this classification.
On the contrary, MyAdChoices provides, for each ad, the probability of error incurred in estimating its class.

Following the same spirit, \cite{Barford14WWW}~presents an ad-crawling infrastructure that does not aim exactly to provide transparency, but to analyze different aspects of advertising at large scale.
Among other aspects, the authors study the average arrival rate of new ads and the distribution of the number of ads and advertisers per page.
In addition, they briefly examine behavioral targeting by following a similar approach to that of~\cite{Carrascosa15CoNEXT,Carrascosa14arXiv}.
They emulate the browsing habits of around 300 users with single-category interests, and try to see
which ads are more targeted to which profiles when visiting a subset of selected Web pages.
Their analysis of profile-targeting assumes that, if an ad is shown more frequently to a given profile than to others, then this ad is targeted to said profile.
Building on certain heuristics, the authors compare the frequency of appearance of each ad (for each profile) with a uniform profile,
and conclude that an ad is targeted if the result of such comparison exceeds a certain threshold.

The proposed framework suffers from the same limitations of the aforementioned two works.
Besides, the authors disregard that ads can be contextual and generic as well, and that the frequency of appearance of ads depends on highly dynamic factors. On the other hand, a practical implementation of this framework on the user side would be unfeasible as it would require that users visit the same pages (to enable the transparency functionality), and exhibit single-category profiles.

A similar platform is proposed in~\cite{LiuHotNet13} that studies the ads delivered to some artificial profiles, in this case built from the AOL search query data set~\cite{AOL06}. The tool is not intended for end-users and provides a framework that aims to study interest-based and contextual advertising at large scale.
The platform, which operates offline and is restricted to DoubleClick ads, analyzes two data sets to this end: the interest categories of \emph{all} ads received both in a tracked session and in an incognito-browsing mode.
The authors then use a binary classifier to decide if an ad belonging to a certain category is interest-based \emph{or} contextual.

The major limitations of this tool come from the simplified and inaccurate model assumed for the ad-delivery process.
First,
it does not take into account generic or untargeted ads.
Secondly, the decision is binary in the sense that the result of an ad classification cannot be contextual \emph{and} interest-based, thus overlooking that the vast majority of ad platforms allow the selection of multiple user-targeting objectives.
Thirdly, such classification relies on the \emph{whole} data set of ads collected in the tracked and incognito sessions, which neglects the fact that DoubleClick (as any ad selector) may construct short-term profiles or use \emph{any} time window to profile users' browsing interests, not necessarily the one that spans the whole browsing history.
Last but not least, the tool in question does not reflect the actual operation of the ad platform it focuses on, namely, that DoubleClick may employ modern RTB technologies to serve ads~\cite{Olejnik14NDSS,Olejnik15PhD}.
On the one hand, the authors seem to assume a baseline scenario, as the user profile is built just from the pages tracked by this ad platform.
But on the other hand, they completely ignore the RTB ad-serving technology,
and the fact that DoubleClick's ad-auction participants may not share the same profiling data.
That is, the authors seem to assume, at the same time, a paranoid scenario, which is contradictory.
We would like to stress that our work addresses all these four issues, by modeling the combination of multiple ad-targeting decisions, relying on the notion of ad selector, building independent user-profile models per ad selector, and considering any possible time window chosen by such entities through the definition of uncertainty class.

Another more recent work for conducting experiments based on artificial profiles is~\cite{Lecuyer14SSYM}, which tracks the personal data collected by several Web services, and tries to correlate data inputs (e.g., e-mails and search queries)
with data outputs (e.g., ads and recommended links).
The proposed platform tackles this correlation problem in a broad sense, and is tested for the ads displayed on Gmail.
The platform relies on the maintenance of a number of \emph{shadow accounts}, that is, replicates of the original account (e.g., an e-mail account), but which differ in a subset of inputs.
All these account instances are operated in parallel by the system and are used to compare the outputs received.
Intuitively, if an ad is displayed more frequently on those accounts sharing a certain input (e.g., an e-mail), and this ad never shows up in the rest of shadow instances, then this input is likely to be the cause of said ad.

The platform in question does not require a shadow account for each possible combination of input data, but a logarithmic number of such accounts in the number of inputs, which makes it suitable for the application where it is instantiated.
However, it would be totally infeasible to extend it so as to analyze the ads received out of this controlled application, for example, while browsing the Web.
First, in terms of scalability. The authors claim to support the correlation of hundreds of inputs (e-mails), with reasonable costs in terms of shadow accounts. This may work for a \emph{single} service provider, but clearly not when considered in the more general context of Web advertising, with thousands of ad companies tracking users throughout the Web~\cite{GhosteryWeb} and around ninety pages visited on average per day~\cite{Nielsen10TR}.
Secondly, creating equivalent shadow browsing profiles on the user side would be impractical in terms of network traffic and computational overhead.
On the other hand, the proposed solution checks which particular input data or combination (with a reduced combination size, to attain the scalability mentioned above) is responsible for a given output data (e.g., an ad). As a result, such platform may work for advertising forms like retargeting, where a single visit may be the cause of an ad display, and for contextual ads, which depend on the page currently being visited. However,
it does not operate on a much coarser granularity level and hence it is not suitable for studying behavioral targeting,
where ads are typically served on the basis of browsing histories accumulated over long time periods.

A couple of refinements of this latter approach are~\cite{Lecuyer15CCS,Datta15PET}, which respectively provide certain statistical validation of its findings and which investigate causation in text-based ads.
The cited works, however, are measurement platforms and suffer from the same limitations in terms of detecting behavioral targeting in a broad sense. Table~\ref{tab:Comparison} summarizes the major conclusions of this section.

\section{Conclusions and Future Work}
\label{sec:Conclusion}
\noindent
In the last few years, as a result of the proliferation of intrusive and invasive ads, the use of ad-blocking and anti-tracking tools have become widespread.
The problem with these technologies is that they pose a binary choice to users
and thus
disregard the crucial role of advertising as the major sustainer of the Internet's free content.

We believe that such technologies are only a short-term solution, and that better tools are necessary to solve this problem in the long term.
Most users are not against ads and are actually willing to accept some ads to help Web sites.
However, this is provided that the ad-delivery process be transparent and users can control the personal information gathered.

Since different users may have different motivations for using ad blockers and anti-trackers,
this paper
proposes a smart Web technology
that can bring transparency to online advertising and help
users enforce their own particular choices over ads.
The primary aim of this technology is, first, to let users know how
their browsing data are exploited by ad companies; and secondly, to enable them to react accordingly by giving them flexible control over advertising.

The proposed technology provides transparency to behavioral targeting by means of two randomized estimators.
The former builds on a theoretical model of the ad-serving process, and capitalizes on the methodology of robust optimization to tackle the problem of modeling the profiles available at ad platforms.
The latter sheds light on these profiles by computing a worst-case uniqueness estimate over all possible profiles constructed by an ad platform.

These two detectors have been integrated into a system architecture that is able to provide ad transparency and blocking services all in real-time, and on the user side.
In terms of transparency, our tool enables users (1) to learn if the ads delivered to them may have been targeted on the basis of their browsing profiles, and (2) to find out whether such profiles may be revealing unique browsing patterns.
In terms of ad blocking, the proposed system allows users to filter out interest-based, non-interested-based and retargeted ads per topic category,
and to specify blocking conditions based on profile uniqueness.

The proposed system has been implemented as a Web-browser extension and assessed in an experiment with 40 participants.
In terms of performance,
the two estimators exhibited running times below 0.5 seconds and reported no errors.
In addition, nearly all pages could be categorized.
We carried out an analysis of behavioral targeting based on the ads and browsing data of those volunteers.
Among other results, our findings show that retargeting is the most common ad-targeting strategy;
that Google's ad companies are the ones leading behavioral and retargeted advertising;
that large firms might be the advertisers mostly delivering profile-based ads;
and that profile uniqueness may not be a widely used criterion to serve ads.

Unlike few previous work on Web transparency, our tool is intended for end-users,
departs from a more faithful, accurate model of the ad-delivery process,
allows for its intricacy and the recently established RTB scheme, and
relies on a mathematically grounded methodology.

Among other aspects, future research should explore possible improvements on the identification and harvesting of ads.
Currently, our extension requires the landing page of an ad to categorize it, but we intend to use optical character recognition technique
to overcome this limitation.
Another strand of future work will investigate enhancements on usability. The proposed tool revisits a small fraction of the pages browsed by the user, and it proceeds by opening a new minimized window in private mode, which might be annoying to some users.

\section*{Acknowledgment}
\noindent
The authors would like to thank Paul Barford, Aaron Cahn and Qiang Ma for helping improve our ad-identification algorithm,
Lukasz Olejnik for his helpful comments, and Mathilde Vernet for helping develop some of the modules used.
This work is partially funded by the Inria Project Lab CAPPRIS.
J. Parra-Arnau is the recipient of a Juan de la Cierva postdoctoral fellowship, FJCI-2014-19703, from the Spanish Ministry of Economy and Competitiveness.

\appendices

\section{Feasibility Problem}
\label{Appendix:Feasibility}
\noindent
This appendix proves the feasibility of the optimization problems~\eqref{eq:Minimax} and~\eqref{eq:uniqueness}.
In particular, it shows that the constraints given by the polyhedron $\mathcal{P}$ are consistent,
or said otherwise, that the set of points satisfying them is nonempty.
For notational simplicity, we rename the tuples $p^{\textnormal{min}}$ and $p^{\textnormal{max}}$
simply with the symbols $r$ and~$s$, respectively.

For~\eqref{eq:Minimax} and~\eqref{eq:uniqueness} to be feasible,
we require
$\sum_i r_i \leqslant 1$ and $\sum_i s_i \geqslant 1$.
To check this, consider the opposite.
On the one hand, having $\sum_i r_i > 1$
and $\sum_i s_i < 1$ leads us to a contradiction, since by definition
$r_i \preceq s_i$.
On the other hand, it is straightforward to verify that, if $\sum_i r_i > 1$, then $\sum_i p_i >1$, and that, if  $\sum_i s_i < 1$, then $\sum_i p_i <1$, which contradict the fact that $p$ is a PMF.

Next, we prove that the requirement $\sum_i s_i \geqslant 1$ is satisfied. The proof of the condition $\sum_i r_i \leqslant 1$ proceeds along the same lines and is omitted.
Recall from Sec.~\ref{sec:System:DetectionIB:Optimal:Robust} that the uncertainty class $\mathcal{P}$
is computed by considering an incremental model on the clickstream.
That is, each time the user visits a Web page,
a new estimate for $p$ is computed from all the pages visited so far.
Then, based on this newly estimated distribution, our system updates $r$ and $s$, if necessary.

The proposed system requires a minimum number of visited pages $w_{\textnormal{min}}$ to estimate $p$.
Following the notation introduced in Sec.~\ref{sec:Implementation:Architecture:Components:Profiling},
we denote by $m_i$ the number of pages that are classified into the topic category $i$.
When such requirement is
met,
the tuples $r$ and $s$ are initialized to
$r_i=s_i = \frac{m_i}{w_{\textnormal{min}}}$
for all $i=1,\ldots,n$.
In other words, $r$ and $s$ become the MLE of $p$.

Let $s_i^m$ be the $i$-th component of the tuple $s$ that results after having visited $m$ pages.
It is immediate to check that
$s_i^{w_{\textnormal{min}}} \leqslant \cdots \leqslant s_i^m$
is a non-decreasing sequence for all $i$,
which implies
that $\sum_i s_i \geqslant 1$.
This proves the feasibility of the problems~\eqref{eq:Minimax} and~\eqref{eq:uniqueness}.

\section{Linear-Program Formulation of the Robust Minimax Detector}
\label{Appendix:LP}
\noindent
Following the methodology developed by~\cite{Boyd04B,Levy08B}, this appendix shows the LP formulation of the robust minimax design problem~\eqref{eq:Minimax}.
From the definitions of $P_i^w$ and $M_{ii}^w$,
it is easy to verify that~\eqref{eq:Minimax} is equivalent to
\begin{equation*}
\max \min_{i=1,2} \inf_{p\in \mathcal{P}} M_{ii},
\end{equation*}
and hence equivalent to the optimization problem
\begin{align}
\quad
\textnormal{maximize}     &\quad\zeta \nonumber \\
\textnormal{subject to}   &\quad\inf\{\tilde{d}^{\,\oT} p : p\in\mathcal{P}\} \geqslant \zeta, \label{eq:appendix:problem1} \\
                                            &\quad 1 - \tilde{d}^{\,\oT} q  \geqslant \zeta, \nonumber\\
                                            &\quad 0\preceq \tilde{d}\preceq\oone. \nonumber
\end{align}
Because the primal problem~\eqref{eq:Minimax} is feasible, Slater's constraint qualification is satisfied and therefore strong duality holds for the Lagrange dual problem associated to the linear program~\eqref{eq:appendix:problem1}.
The dual problem in question is
\begin{align*}
\quad
\textnormal{maximize}     &\quad\mu^{\oT}p^{\textnormal{min}} - \lambda^{\oT} p^{\textnormal{max}} + \nu  \\
\textnormal{subject to}   &\quad\mu - \lambda + \nu\oone \preceq \tilde{d},\\
                                            & \quad \lambda\succeq 0,\mu \succeq 0,
\end{align*}
where $\lambda,\mu,\nu$ are the Lagrange multiplier vectors associated with the minimization problem~\eqref{eq:appendix:problem1},
and $p^{\textnormal{min}},p^{\textnormal{max}}$ determine the polyhedron $\mathcal{P}$ defined in~\eqref{eq:unclass}.
Leveraging on this dual problem, we immediately derive the LP formulation~\eqref{eq:MinimaxLP}.

\bibliographystyle{IEEEtran}

\end{document}